\documentclass[aps,pre,showpacs,superscriptaddress,twocolumn,amssymb,floats,epsfig]{revtex4-1}

\usepackage{graphicx}
\usepackage{epsfig}
\usepackage{amsfonts}
\usepackage{amsmath}
\usepackage{amssymb}
\usepackage{subfigure}
\usepackage{color}
\usepackage{soul}

\newcommand{\ct}{\cite}
\newcommand{\bi}{\bibitem}
\newcommand{\be}{\begin{equation}}
\newcommand{\ee}{\end{equation}}
\newcommand{\ba}{\begin{eqnarray}}
\newcommand{\ea}{\end{eqnarray}}

\newcommand{\ket}[1]{|#1\rangle}

\newcommand{\cmp}
{\affiliation{Condensed Matter Physics Division,
Saha Institute of Nuclear Physics, 1/AF Bidhannagar, Kolkata 700064, India.}}
\newcommand{\barasat}
{\affiliation{Barasat Government College, Barasat, Kolkata 700124, India.}}
\newcommand{\germany}
{\affiliation{Max Planck Institute for the Physics of Complex Systems, N\"othnitzer Str. 38, Dresden 01187, Germany}}

\begin{document}

\title{Dynamics of decoherence of an entangled pair of qubits locally connected to a one-dimensional 
disordered spin chain }

 \author{Sudip Mukherjee}
 \email{sudip.mukherjee@saha.ac.in}
  \barasat \cmp
 \author{Tanay Nag}
 \email{tanay@pks.mpg.de }
 \germany
\begin{abstract}

{We study the non-equilibrium evolution of concurrence of a Bell pair constituted of two 
qubits, through the measurement of Loschmidt echo (LE) under the scope of generalized central spin 
model. The qubits are locally coupled to a one dimensional disordered Ising spin chain. We first 
show that in equilibrium situation the derivative of LE is able to detect the extent of Griffiths 
phase that appeared in presence of disordered transverse field only. While in the non-equilibrium 
situation, the spin chain requires a temporal window to realize the effect of disorder. We show 
that within this timescale, LE falls off exponentially and this decay is maximally controlled by 
the initial states and coupling strength. Our detail investigation suggests that there actually 
exist three types of exponential decay, a Gaussian decay in ultra short time scale followed by 
two exponential decay in the intermediate time with two different decay exponents. The effect of 
the disorder starts appearing in the late time power law fall of LE where the power law exponent 
is strongly dependent on disorder strength and the final state but almost independent of initial 
states and coupling strength. This feature allows us to indicate the presence of Griffiths phase. 
To be precise, continuously varying critical exponent and wide distribution of relaxation time 
imprint their effect in LE in the late time limit where the power law fall is absent for quenching 
to a Griffiths phase. Here, LE vanishes following the fast exponential fall. Interestingly, for 
off-critical quenching LE attains a higher saturation value for increasing disorder strength, 
otherwise vanishes for a clean spin chain, referring to the fact that disorder prohibits the rapid 
decay of entanglement in long time limit. Moreover, we show that disorder is also able to destroy 
the light cone like nature of post quench quasi-particles as LE does not sense the singular time 
scales appearing for clean spin chain with qubits coupled at symmetric positions.}

\end{abstract}
\maketitle
\section{Introduction}
In recent year, there has been an upsurge of studies in various quantum information theoretic 
measures such as fidelity \ct{gu08}, decoherence \ct{quan06,damski11a}, concurrence \ct{wootters01,horodecki01}, 
quantum discord \ct{ollivier01} and entanglement entropy \ct{vidal03} connecting the quantum 
information science \ct{kitaev09, nag12b, sachdeva14,suzuki15, nag15, nag16, rajak16}, statistical 
physics and condensed matter physics \ct{chakrabarti96, sachdev99, polkovnikov11, dutta15} in a 
concrete way. In particular, the effect of quantum criticality appears in the ground-state correlation
that becomes maximum at the quantum critical point (QCP); for example, the concurrence, a separability 
based approach to measure the quantum correlation, can detect as well as characterize a QCP. Furthermore, 
considering a central spin model, a single qubit globally coupled to an environmental spin chain that 
exhibits a quantum phase transition, a plethora of the studies is devoted to investigate the decoherence 
of the qubit by probing the Loschmidt echo (LE) \ct{cucchietti05,quan06,yuan07}; LE is defined as the 
square of overlap of the two wave functions evolved with two different Hamiltonians while initially both 
the states are prepared in the ground state of the one of the Hamiltonians. It has been experimentally 
demonstrated using NMR quantum simulator that the LE shows a dip at the QCP of a finite antiferromagnetic 
Ising spin chain, hence it suffices as an ideal detector of a QCP \ct{zhang09}. 

Motivated by the seminal work on random transverse field Ising model \ct{fisher92} and random Heisenberg 
antiferromagnetic chains \ct{dasgupta80} using the strong-disorder renormalization group technique, the 
disordered spin systems have grabbed an enormous attention due to its non-trivial modifications on the phase 
diagram obtained in the clean disorder free case \ct{igloi05, mckenzie96, bunder99}; in the later works, the 
critical behavior is investigated using a mapping to random-mass Dirac equations. Interestingly, randomness 
might lead to Griffiths phases  and new universality classes \ct{griffiths69}. It is noteworthy that the 
emergence of Griffiths phase is investigated using the finite size scaling of fidelity susceptibility and 
its distribution probability \ct{garnerone09}. Similarly, in the context of entanglement entropy, it has 
been shown for the disordered spin chain that an effective central charge dictates its behavior \ct{peschel05, refael07}. 
The critical properties of long range disordered transverse Ising model \ct{juhasz14} and associated many 
body localization transitions \ct{li16} are also extensively studied in equilibrium.

Simultaneously, slow quenching dynamics of many body quantum systems \ct{zurek05} and the quantum information 
theoretic measures \ct{sengupta09} emerge as interdisciplinary fields of research \ct{nag13}. One of the other 
ways of generating such a non-equilibrium dynamics is a sudden quench \ct{dutta15}. The sudden quench dynamics 
of entanglement entropy is  investigated for disordered quantum spin chains \ct{igloi12}. Recently, entanglement 
entropy in the context of many body localization transition is examined using disordered Ising chain following 
this quench \ct{Kjall14}. In parallel, the relaxation dynamics and thermalization after a quantum quench in 
disordered chain has also captured attention \ct{ziraldo12}. Focusing on our work, the model considered here is 
referred as the generalized central spin model (GCSM) where two qubits are locally coupled to the environmental 
spin chain. It noteworthy that the sudden quench is also employed in the environmental clean spin chain to study 
the generation of concurrence \ct {nag16} and decay of concurrence \ct{wendenbaum14} between a pair of  qubits 
considering GCSM.

Given the recent studies on disentanglement of a pair of qubits coupled to clean environment, our main aim here is 
to investigate the effect of disorder  in the environmental chain on the non-equilibrium evolution of concurrence 
which is directly estimated from LE. The temporal characterization of LE and subsequently, concurrence
for a GCSM with disordered transverse Ising  chain to the best of our knowledge is completely new. In the process, by computing 
the derivative of LE we first exemplify a situation where randomness present only in transverse field leads to a Griffiths 
phase. Secondly, The signature of Griffiths phase is also imprinted in the temporal evolution of LE; long time power law decay 
of LE is absent for quenching inside Griffiths phase. Although, a short time rapid exponential fall is observed for all
types of quenching schemes. Moreover, our study suggests that disorder helps in preserving entanglement between two 
initially entangled qubits. Furthermore, we find that the ultra-short time Gaussian fall of LE remains unaltered even 
in the presence disorder. Our investigation suggests that equilibrium characteristics and non-equilibrium 
dynamics of LE might serve as useful indicator for detecting Griffiths phase. Finally, we show that the  quasi-particles 
generated after the quench do not propagate in the light cone like manner after a threshold disorder strength.

This paper is organized as follows: In Sec.~\ref{model} we introduce the CGCM consisting of two qubits locally connected
to two sites of an  Ising chain in presence of disordered transverse field. We define the concurrence. It is derived from 
the $4\times 4$ reduced density matrix of the two qubits obtained by tracing out the environmental degrees of freedom; In 
Sec.~\ref{result}, we illustrate our results for weak as well as strong  coupling case and analyze the ultra-short, short 
and long time behavior of concurrence specifically, LE. Finally, we provide our concluding remarks in Sec.~\ref{conclusion}.
\section{model}
\label{model}
We consider GCSM, comprised of two qubits (system) and an environmental spin chain, to study the entanglement dynamics between 
these two qubits when the environment is suddenly quenched. The qubits are noninteracting and they are locally coupled with the 
environment. The Hamiltonian of the total 
system (qubits $+$ environment) is given by
\begin{equation}
H_{T} = H_{EN} + H_{Q}. 
\end{equation}
Here $H_{EN}$ is the Hamiltonian of the environment, in our case which is a one dimensional chain of $N$ Ising spins in presence
of random transverse field. The interaction Hamiltonian between the qubits and the  environment is represented as  $H_{Q}$. This 
model is widely studied to measure decoherence and entanglement between qubits in the qubit-bath set up\ct{cormick08}.

The Hamiltonian of the environment is given by
\begin{equation}
H_{EN} = -J\sum_{i=1}^N {\sigma}_{i}^{x}{\sigma}_{i+1}^{x} - \sum_{i=1}^{N}{\Gamma}_i{\sigma}_{i}^{z}, 
\label{H_en}
\end{equation}
where $\sigma_i^x$, $\sigma_i^z$ are the $x$ and $z$ components of Pauli spin matrices respectively and ${\Gamma}_i$ is the 
transverse field at the $i$-th site. Such ${\Gamma}_i$s are distributed following Gaussian distribution 
$P({\Gamma}_i) =(1/\sqrt{2\pi{\Omega}^2})\exp(-({\Gamma}_i - \Gamma)^2/2{\Omega}^2)$, where $\Gamma$ is the mean and $\Omega$ 
is the standard deviation of the distribution. The standard deviation $\Omega$ reveals the strength of disorder in the values 
of transverse field. Hereafter the average or mean value of the transverse field will be simply referred as transverse field.
The nearest neighbor spin-spin interaction is denoted by $J$ and we consider periodic boundary condition.

Let us revisit the critical behavior of disordered Ising chain quickly; one can consider $J_i$ and $\Gamma_i$ to be random 
interaction and site dependent random coupling. The  spin chain becomes critical when the average value of the field matches 
with the average value of the coupling. It has been shown using the strong disorder renormalization group that, at the QCP, 
the time scale $\hat t$ and length scale $N$ are related by $\log_e(\hat t) \sim N^{1/2}$. As a result, dynamical exponent
$z$ at criticality acquires an infinite value.  Another interesting feature  is the generation of Griffiths phase in the 
vicinity of the QCP; the distribution of relaxation times becomes broadened due to Griffiths singularities. The Griffiths 
phase is specifically characterized by a dynamical exponent $z$ which depends on the distance ($\delta_G$) from the  QCP 
($\delta_G= 0$) of the pure spin chain. The width of the Griffiths phase i.e., $\delta_G$ has been analytically obtained using 
Dirac-type equation with random mass in the continuum limit \ct{mckenzie96}; this also confirms the numerical finding obtained 
using real space renormalization group  decimation  technique \ct{fisher92}. We note that this Griffiths phase appears in the 
disordered side of the phase diagram. We here consider homogeneous interaction strength, $J_i=J$; this belongs to another 
prototypical model for disordered quantum spin chain which falls in the universality class of Ising  transitions for all 
values of $J_i$.

We consider the qubits are coupled at the sites $1$ (qubit $A$) and $1+d$ (qubit $B$) of the environmental spin chain given 
in (\ref{H_en}). The interaction Hamiltonian of the qubits $H_{Q}$ is given by 
\begin{equation}
H_{Q} = - \Delta (\lvert\uparrow \rangle \langle \uparrow\rvert_{\text{A}} \otimes {\sigma}_{1}^{z} +
\lvert\uparrow \rangle \langle \uparrow\rvert_{\text{B}} \otimes {\sigma}_{1+d}^{z}).
\end{equation}
Here $\lvert\uparrow \rangle$ is the eigenstate of ${\sigma}_{\text{A,B}}^{z}$ such that 
${\sigma}_{\text{A,B}}^{z}\lvert \uparrow \rangle = \lvert\uparrow \rangle$. The coupling strength is denoted by 
$\Delta (> 0)$. ${\sigma}_{i}^{z}$ represents the environmental spin at site $i$.

The initial state of the qubits is a maximally entangled Bell state: $\lvert\phi \rangle_{\text{A,B}} = \frac{1}{\sqrt{2}}
(\lvert\uparrow_A \uparrow_B \rangle + \lvert\downarrow_A \downarrow_B \rangle)$. The environment is assumed to be in its 
ground state $|\Phi(\Gamma^{I})\rangle_g$ where $\Gamma^I$ is the initial value of the transverse field. Therefore, the 
initial state of the total system is $|\psi(0)\rangle = |\phi_{\text{A,B}}\rangle\otimes|\Phi(\Gamma^I)\rangle_g$. At initial 
time $t=0^+$, we suddenly quench the transverse field from $\Gamma^I$ to a final value $\Gamma^F$. Such quenching results in a 
non-equilibrium time evolution of the composite system by changing the instantaneous state of the environment. The time evolution 
of the environment depends on the initial state of the qubits. If the qubits are in the state $\lvert\downarrow_A \downarrow_B\rangle$ 
then the spin chain will evolve with the Hamiltonian $H_{\downdownarrows} = H_{EN}(\Gamma^F)$. On the other hand, if the qubits
are in the state $\lvert\uparrow_A \uparrow_B\rangle$ then the time evolution of the environment will be governed by the Hamiltonian 
$H_{\upuparrows} = H_{EN}(\Gamma^F) - \Delta(\sigma_{1}^z + \sigma_{1+d}^z)$. The state of the total system at instant of time 
$t$ can be written as
\be
|\psi(t)\rangle = \frac{1}{\sqrt{2}}[\lvert\upuparrows\rangle\otimes|\Pi_{\upuparrows}(t)\rangle_{EN} + 
\lvert\downdownarrows\rangle\otimes|\Pi_{\downdownarrows}(t)\rangle_{EN}].
\label{eq_psit}
\ee
where $\lvert\Pi_{\lambda}(t)\rangle_{EN} = e^{-i H_{\lambda}(\Gamma^F)t}\lvert\Phi(\Gamma^I)\rangle_g $ with 
$\lambda = \upuparrows, \downdownarrows $. Here we omit $A$ and $B$ subscripts for simplicity.

In order to measure the decoherence of the qubits, we compute LE $L$ which is defined as
\begin{equation}
L_{\upuparrows,\downdownarrows} = {{\lvert}{}_{g}{\langle}\Phi(\Gamma^I)\rvert e^{i H_{\downdownarrows}(\Gamma^F)t}e^{-i H_{\upuparrows}(\Gamma^F)t}\lvert\Phi(\Gamma^I)\rangle_g\rvert}^2.
\label{le1}
\end{equation}
The dynamics of the LE is governed by the environmental Hamiltonians $H_{\upuparrows}$ and $H_{\downdownarrows}$.

One can represent both the Hamiltonians by the Fermionic operators using Jordan-Wigner transformation 
$\sigma_{j}^{+} = e^{i\pi\sum_{l=1}^{j-1}{\sigma}_{l}^{\dagger}{\sigma}_{l}}c_{j}^{\dagger}$, 
$\sigma_{j}^{-} = c_{j}e^{-i\pi\sum_{l=1}^{j-1}{\sigma}_{l}^{\dagger}{\sigma}_{l}}$ where 
$\sigma_{j}^{+}$ and $\sigma_{j}^{-}$ are define as $\sigma_{j}^{\pm} = \frac{\sigma_{j}^{x} \pm i\sigma_{j}^{y}}{2}$. 
The generic form of the environment Hamiltonian can be written in terms of $c^{\dagger}$, $c$  given by 
\begin{equation}
H = \sum_{i,j} [c_{i}^{\dagger}A_{ij}c_j + \frac{1}{2}(c_{i}^{\dagger}B_{ij}c_j^{\dagger} + H.c)].
\end{equation}
Here $A$ and $B$ are two $N \times N$ matrices with $A_{ij} = -2\Gamma^I\delta_{ij} - J(\delta_{j,i-1} + \delta_{i,j-1})$, 
$B_{ij} = J(\delta_{j,i+1} - \delta_{i,j+1})$. The Hamiltonian $H$ can be written as 
\begin{equation}
H = \frac{1}{2}\varPsi^{\dagger}\mathcal{H}\varPsi~~\text{where}~~
\mathcal{H}=
\left(
\begin{array}{cc}
-A & -B\\B & A
\end{array}
\right)
\label{ham1}
\end{equation}
Here $\varPsi^{\dagger} = (\bf{\mathcal{C}},\bf{\mathcal{C}^{\dagger}})$ $= (c_1, c_2,...,c_N,c_1^{\dagger}, c_2^{\dagger},....,c_N^{\dagger})$.
The Hamiltonian $\mathcal{H}$ (\ref{ham1}) can be diagonalized by the unitary matrix $U$ given by  
\begin{equation}
U=
\left(
\begin{array}{cc}
g & h\\h & g
\end{array}
\right).
\label{u_mat}
\end{equation}

One  can now write LE (\ref{le1}) in terms of covariant matrices \ct{wendenbaum14}
$\mathcal{L}_{\lambda}(t) = {}_{EN}{\langle}\Pi_{\lambda}(t)\lvert\varPsi\varPsi^{\dagger}\lvert\Pi_{\lambda}(t)\rangle_{EN}$  given by 
\begin{equation}
L_{\upuparrows,\downdownarrows} = {\lvert {\rm det}(\mathcal{I} - \mathcal{L}_{\downdownarrows}(t) - \mathcal{L}_{\upuparrows}(t))\rvert}^{1/2} 
\label{echo1}
\end{equation}
The time evolution of $\mathcal{L}$ is written as $\mathcal{L}_{\lambda}(t) = e^{-it\mathcal{H}_{\lambda}}\mathcal{L}(0)e^{it\mathcal{H}_{\lambda}}$. 
Such covariant matrix at $t=0$, can be represented as 
\begin{equation}
\mathcal{L}(0)=
\left(
\begin{array}{cc}
\langle\bf{\mathcal{C}^{\dagger}}\bf{\mathcal{C}}\rangle & \langle\bf{\mathcal{C}^{\dagger}}\bf{\mathcal{C}^{\dagger}}\rangle\\
\langle\bf{\mathcal{C}}\bf{\mathcal{C}}\rangle & \langle\bf{\mathcal{C}}\bf{\mathcal{C}^{\dagger}}\rangle
\end{array}
\right)=
\left(
\begin{array}{cc}
h^{T}h & h^{T}g\\g^{T}h & g^{T}g
\end{array}
\right)
\end{equation}
Here $\langle . \rangle$ denotes the expectation value in the initial ground state of the environment $\lvert\Phi(\Gamma^I)\rangle_g$.
We note that the method of computing LE is directly borrowed from the Ref.\ct{wendenbaum14} where the environment is considered to 
be a clean spin chain. We give a derivation of the above Eq.~(\ref{echo1}) in the Appendix.

Having computed the LE, now we can calculate the concurrence between the qubits from the reduced density matrix for the qubits.
This reduced density matrix of the qubits can be constructed by tracing out the environmental degrees of freedom from the composite 
density matrix obtained from $\ket{\psi(t)}$ given in Eq.~(\ref{eq_psit}).  In the basis 
{$\{\ket{\uparrow\uparrow},\ket{\uparrow\downarrow},\ket{\downarrow\uparrow},\ket{\downarrow\downarrow}\}$}, 
the reduced density matrix for the two qubits system is given by
\be \rho_s(t) = \frac {1}{4} \left[ \begin{array}{cccc} 
1 & 0& 0& d_{\uparrow \uparrow, \downarrow \downarrow} \\
0 & 1 & 0& 0\\
0& 0& 1 &0\\
d^*_{\uparrow \uparrow, \downarrow \downarrow} & 0& 0 & 1 
\end{array} \right], \label{eq:rho_s} \ee
where $d_{\uparrow\uparrow,\downarrow\downarrow}={}_{EN}\langle \Pi_{\uparrow\uparrow}(t)|\Pi_{\downarrow\downarrow}(t)\rangle_{EN}$.
The LE corresponding to different channels is  thus given by
 $L_{\uparrow \uparrow,\downarrow\downarrow}(t)=|d_{\uparrow \uparrow,\downarrow\downarrow}(t)|^2$ and  its explicit form is written in Eq.~ (\ref{le1}).

Now, using the density matrix $\rho_s(t)$ in Eq.~(\ref{eq:rho_s}), one can calculate the concurrence between the two qubits. 
The concurrence is given by 
\be
 {C(\rho_s)=\rm max(0,\sqrt{\epsilon_1}-\sqrt{\epsilon_2}-\sqrt{\epsilon_3}-\sqrt{\epsilon_4})},
\label{eq:cnc}
\ee
where {$\epsilon_i$'s are the  eigenvalues} in a descending  order of the 
non-Hermitian matrix $M=\rho_s\hat\rho_s$ with $\hat \rho_s$ defined as 
\be
\hat\rho_s=(\sigma^y\otimes\sigma^y)\rho_s^*(\sigma^y\otimes\sigma^y).
\label{eq:R_mat}
\ee
The eigenvalues are coming to be $\epsilon_1(t)=1/4(1+|d_{\uparrow \uparrow,\downarrow\downarrow}(t)|)^2$,
$\epsilon_2(t)=1/4(1-|d_{\uparrow \uparrow,\downarrow\downarrow}(t)|)^2$, $\epsilon_{3,4}=0$.
One can easily show that the concurrence becomes $C(\rho_s)=\sqrt{L_{\uparrow \uparrow,\downarrow\downarrow}(t)}=
|d_{\uparrow \uparrow,\downarrow\downarrow}(t)|$. Therefore, the LE of the environment coupled to the system  is the main
quantity of interest which measures the entanglement between the qubit during the time evolution. We shall focus only on 
the LE in the subsequent result section.

\section{Numerical results}
\label{result}

\subsection{Existence of Griffith phase}

The Loschmidt echo can act as an indicator of quantum phase transition in the equilibrium \ct{quan06}. This is also true 
for non-equilibrium situation where derivative of LE with respect to a parameter of the quantum Hamiltonian shows a singular 
behavior at QCP \ct{wendenbaum14}. The question we ask here is that whether for a disordered chain the derivative of LE 
still shows a peak around the QCP of the clean chain. Our study suggests that first derivative of $L$  with respect to the 
transverse field shows a peak around the boundary of the Griffiths phase that is created in presence of disorder. The position 
of such peak does not coincide with the peak appearing at the QCP of the clean spin chain.

\begin{figure}
\begin{center}
\includegraphics[width=3.80cm]{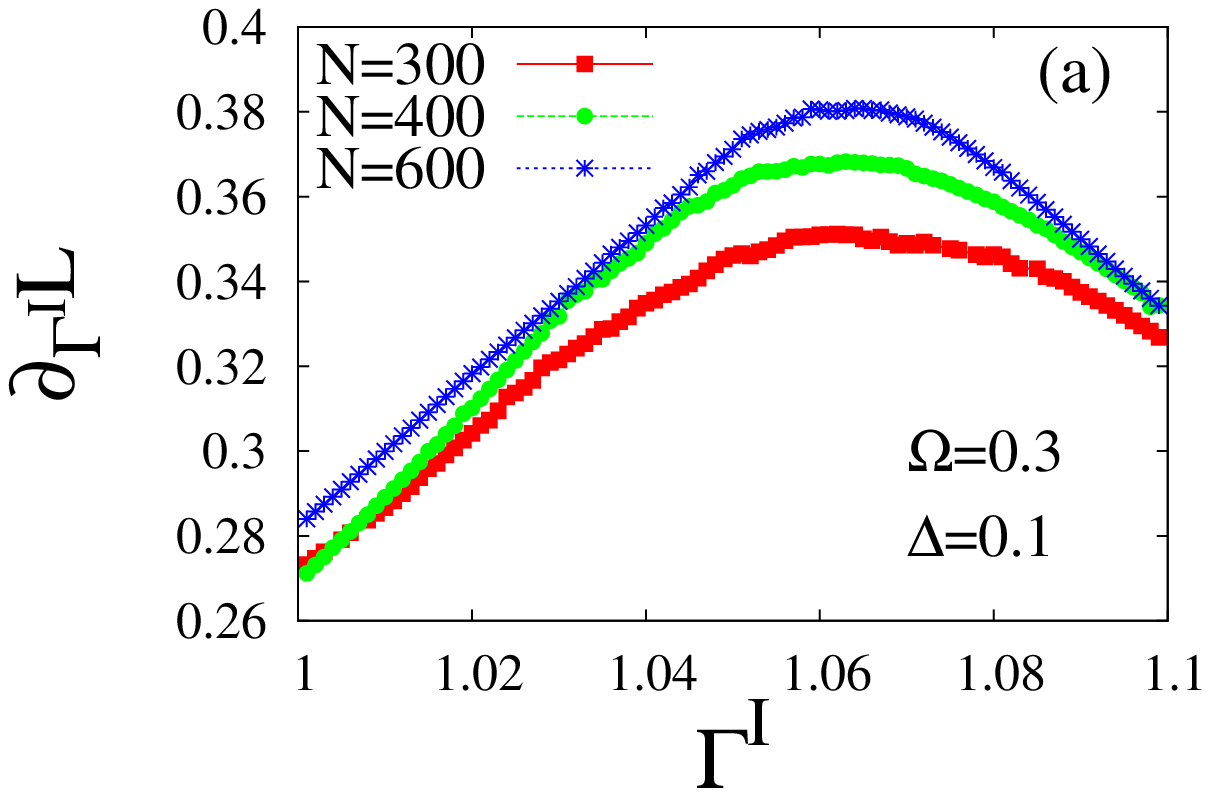}
\includegraphics[width=3.80cm]{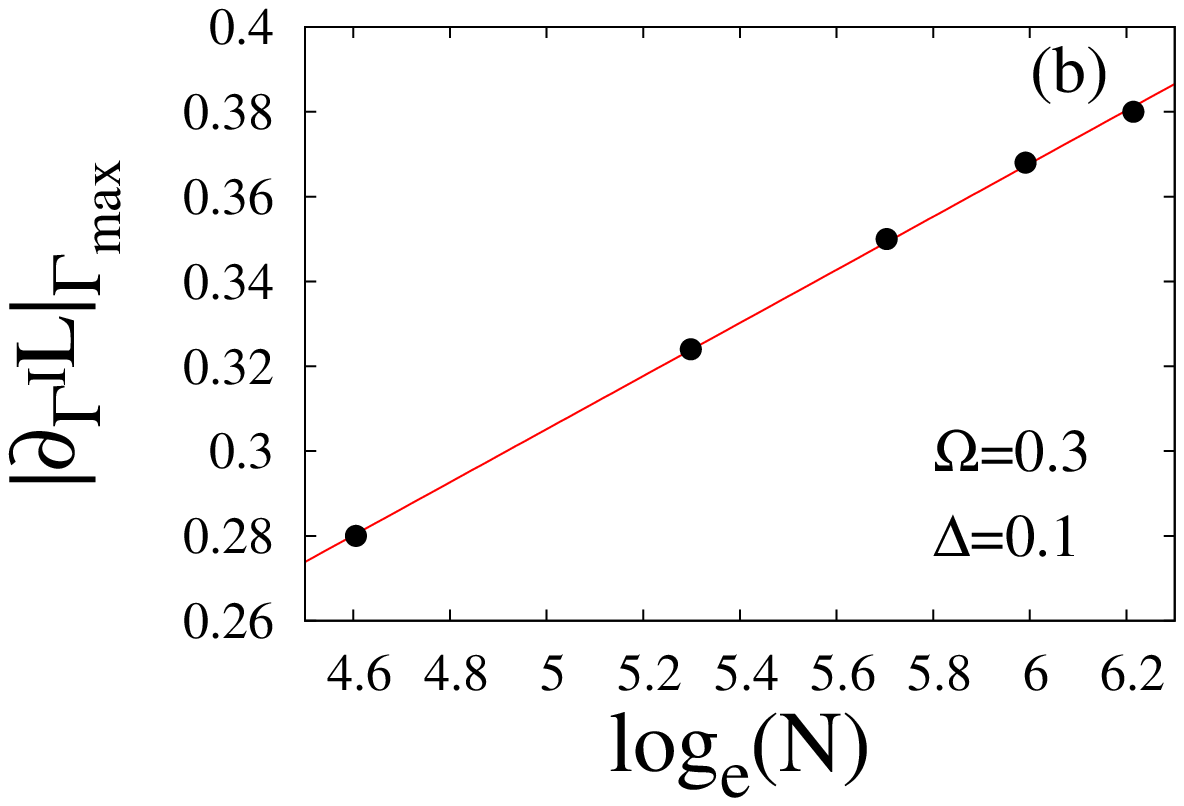}
\end{center}
\caption{(Color online) (a) The variations of the derivative of Loschmidt echo (LE) $\partial_{\Gamma^I}L$ with the initial 
transverse field $\Gamma^I$ for different system sizes $N$ are shown. Here the coupling strength $\Delta = 0.10$ and the 
disorder strength is $\Omega = 0.30$. For all the system sizes, $\partial_{\Gamma^I}L$ shows a peak at $\Gamma^I \approx 1.06$ 
which reveals the existence of Griffith phase in the disordered Ising chain. Here, we set time $t = 10$ and the final value of 
transverse filed is $\Gamma^F = 1.50$. (b) The peak height of the derivative grows linearly with $\log_e(N)$}. 
\label{Griffith_phase}
\end{figure}

\begin{figure*}[htb]
\begin{center}
\includegraphics[width=6.0cm]{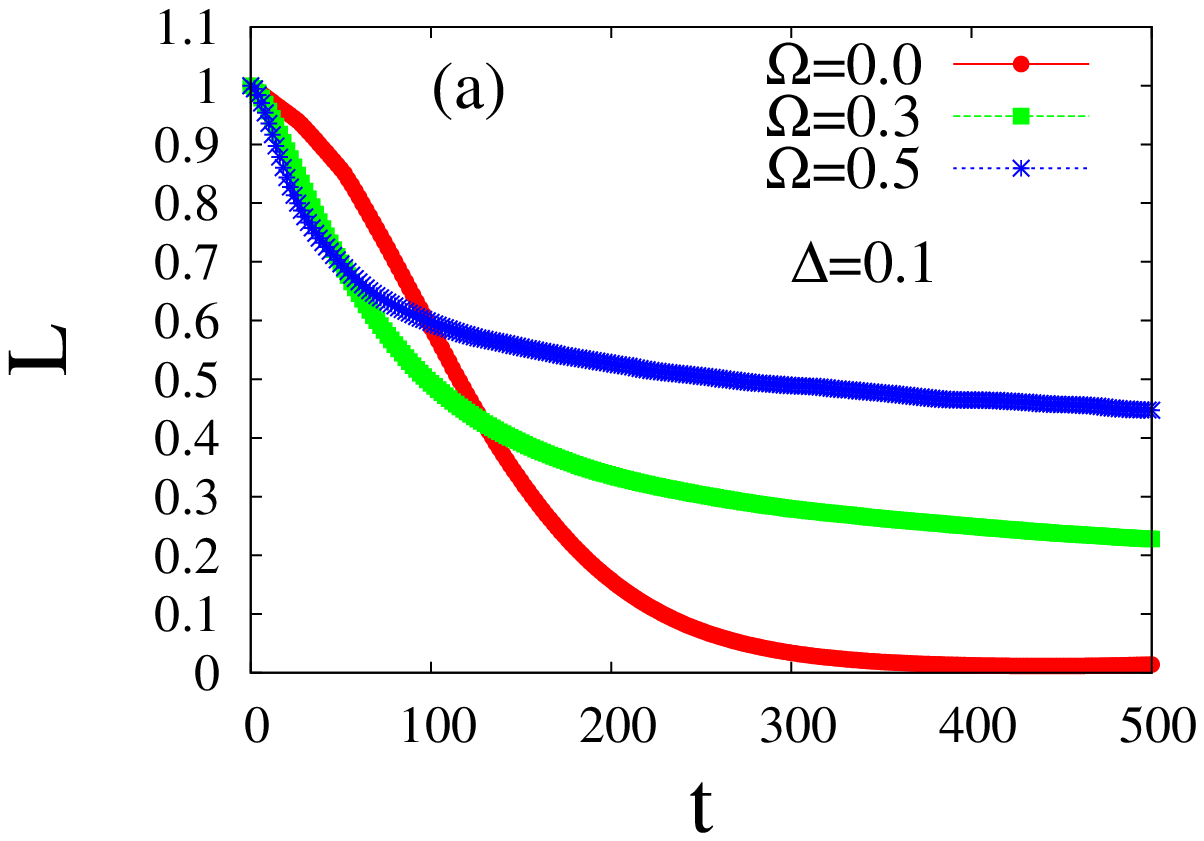}
\includegraphics[width=6.0cm]{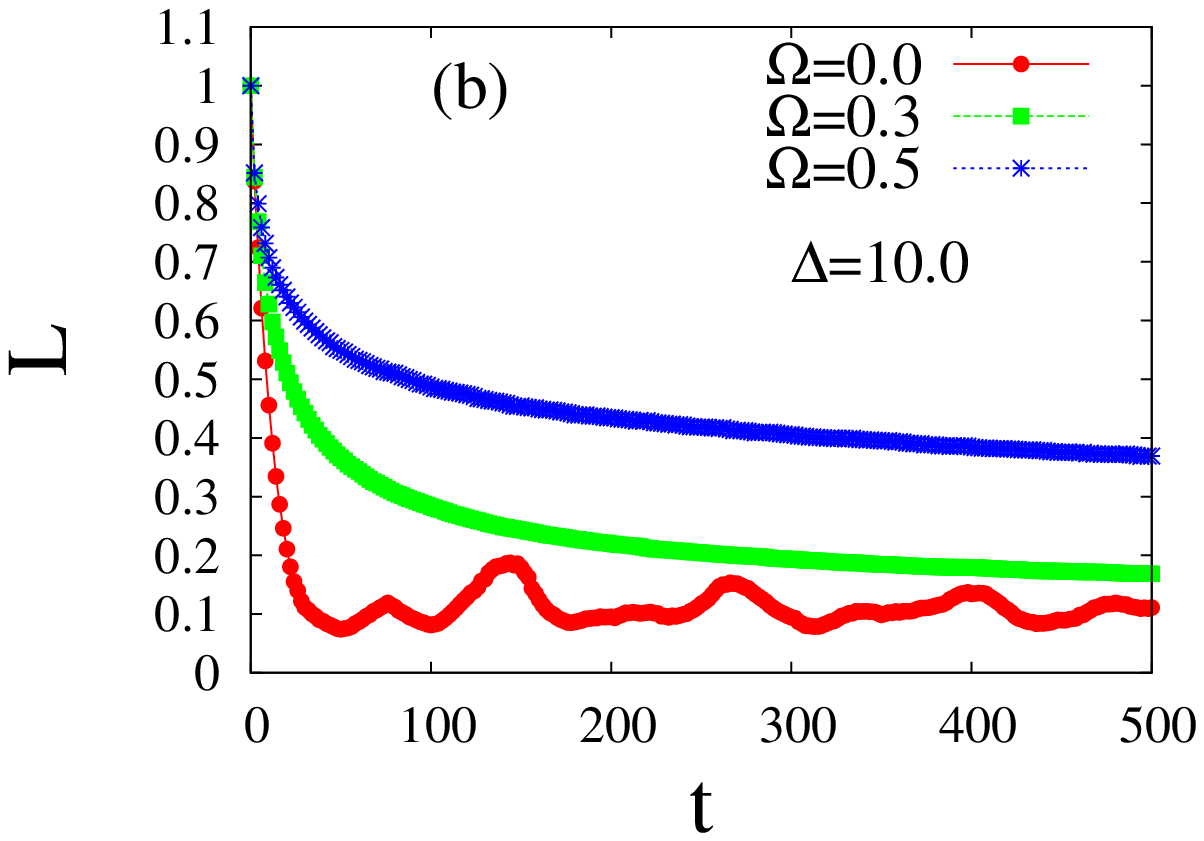}
\includegraphics[width=6.0cm]{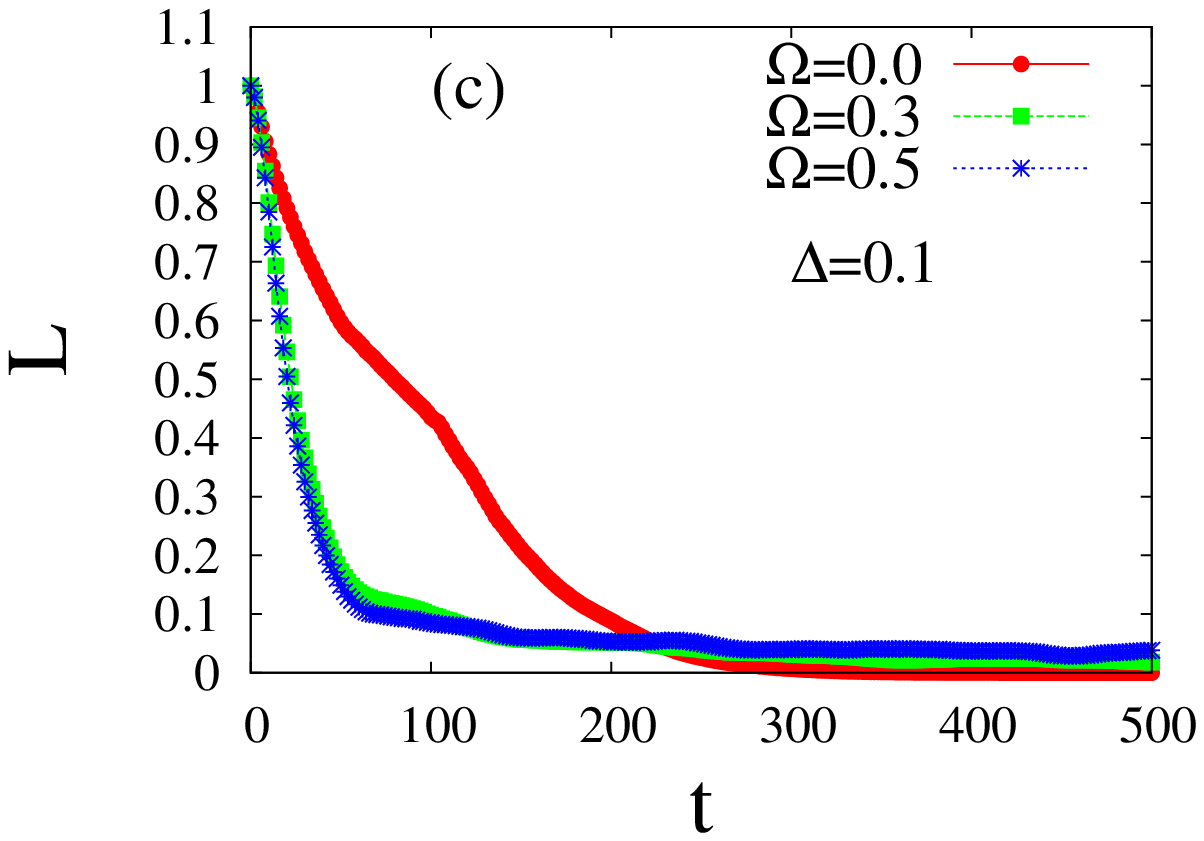}
\includegraphics[width=6.0cm]{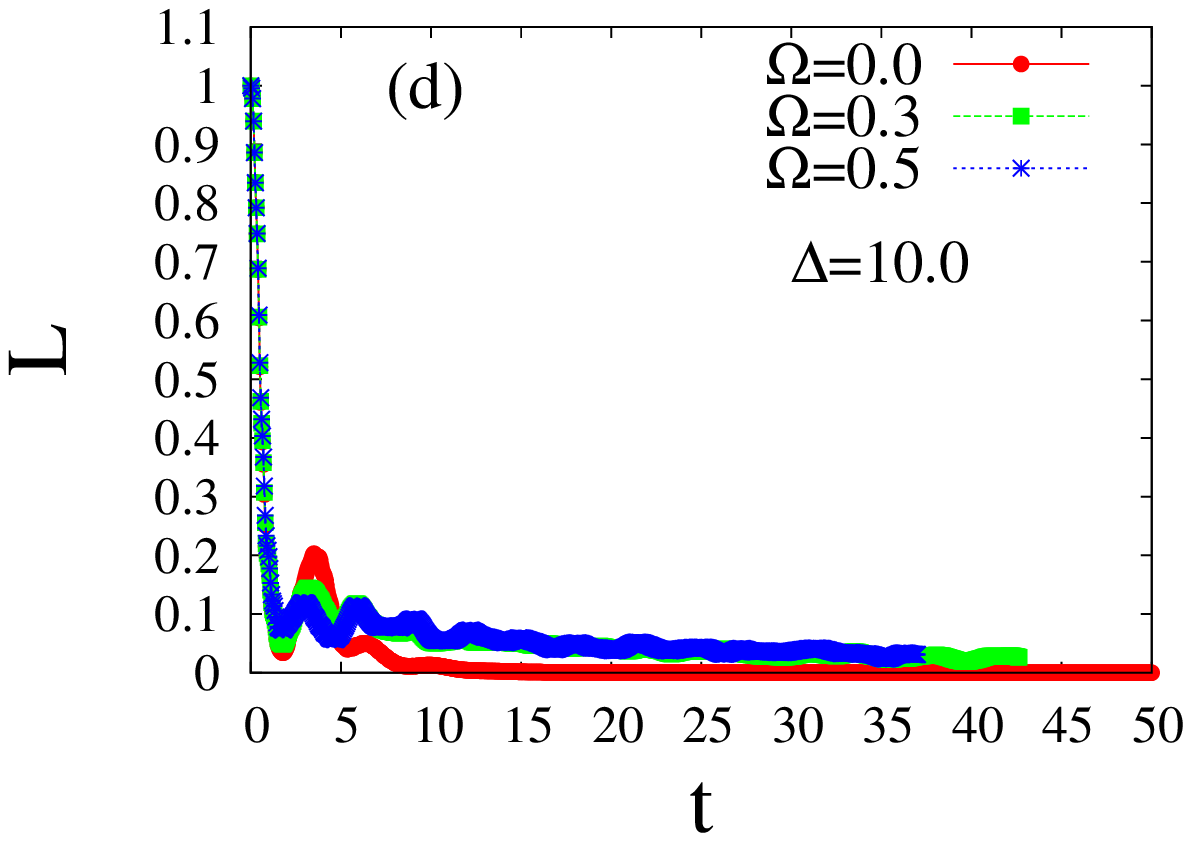}
\end{center}
\caption{(Color online) The temporal dynamics of LE  are shown for different types of quenching schemes. In (a) and (b), with 
$\Delta=0.1$ and $10.0$ respectively, the time evolution of $L$ are plotted for different values of disorder strengths 
$\Omega = 0.0, 0.3, 0.5$. Here the transverse field is quenched from $\Gamma^I=10.0$ to $\Gamma^F=2.0$ (i.e., quenching within 
same paramagnetic phase). In (c) and (d), with $\Delta=0.1$ and $10.0$ respectively, the time evolution of $L$ are plotted for 
the same set of disorder strengths; the transverse field is quenched from $\Gamma^I=10.0$ to $\Gamma^F=0.5$ (i.e., quenching from 
paramagnetic to ferromagnetic phase). We consider system size to be $N=100$.}
\label{para_para}
\end{figure*}

We numerically calculate the derivative of Loschmidt echo ($\partial_{\Gamma^I}L$) for different values of initial 
transverse field $\Gamma^I$  by keeping final value of transverse field fixed at $\Gamma^F=1.5$  and time is fixed 
at $t = 10$. (see Fig.\ref{Griffith_phase}a). The values of coupling strength and coupling distance are $\Delta  = 0.10$ and $d = 1$ 
respectively. In our calculation, we take disorder strength $\Omega = 0.30$ and the disorder average is made over 
$500$ configurations. The derivative of LE shows peak at $\Gamma^I \approx 1.06$ which does not show any tendency 
to shift towards  clean critical point ($\Gamma^I = 1.00$)  when the system size increases from $N = 300$ to $N= 600$. 
This shift of the critical point indicates the existence of the Griffith phase in the disordered Ising spin chain; this 
phase arises due to the inhomogeneity  of the transverse field. The derivative shows a peak at the boundary between the 
Griffiths phase and the disordered phase. Therefore, our numerical study indicates that the Griffiths phase is extended 
from the $\Gamma = 1.0$ to $\Gamma = 1.06$. The analytical expression of the Griffiths phase for disorder present in both 
interaction and transverse field \ct{young96} might not be applicable for our case as we are dealing with randomness only 
in transverse field. {In our case, we think that due to the disordered transverse field, the competition between 
spin-spin interaction and quantum fluctuation (i.e., local transverse field) are not identical in the individual sites. 
As result of that there exist local order  in the spin chain and the thermodynamic observables essentially 
become non self-averaging. In this way one would expect that Griffiths phase  emerges in presence of disorder transverse field.}
Moreover, we note that if ${\Gamma}_i$s are distributed with only two possible values ${\Gamma}_i = \pm \Gamma$, then one would 
not expect to see  any Griffiths phase. We also note that using fidelity susceptibility  the boundary of Griffiths phase has also 
been probed \ct{garnerone09}. Similar to the clean system~\ct{wendenbaum14}, here also the peak height logarithmically increases 
with the system size i.e., $|\partial_{\Gamma^I}L|_{\Gamma_{\rm max}} \sim \log_e(N)$ (see Fig.\ref{Griffith_phase}b).

\subsection{Quenching dynamics of the spin chain}
We numerically study the quenching dynamics of concurrence, determined by LE $L$, between two qubits a disordered transverse
field  Ising chain of length $N = 100$. We note that in all the subsequent numerical calculation, the disorder average is made 
over $500$ configurations of the environmental chain and the value of coupling distance is $d = 1$. The nature of the coupling 
is determined by comparing it with the bare energy gap of the environment i.e., $2$ set by the  clean spin chain with periodic 
boundary condition. We investigate the GCSM for strong coupling strength $\Delta>2$ as well as weak coupling limit i.e., $\Delta<2$ 
following ferromagnetic and paramagnetic quenches in the environmental spin chain. Furthermore, we extend our study of non-equilibrium 
dynamics of LE following a quench inside the Griffiths phase. We analyze the dynamics for different values of disorder strengths 
along with the clean limit $\Omega=0.0$. Our focus is to extensively investigate the temporal decay of LE under these different 
quenching schemes; we hence study the ultra-short, short and long time domains in the non-equilibrium evolution of LE.

\begin{figure*}
\begin{center}
\includegraphics[width=5.5cm]{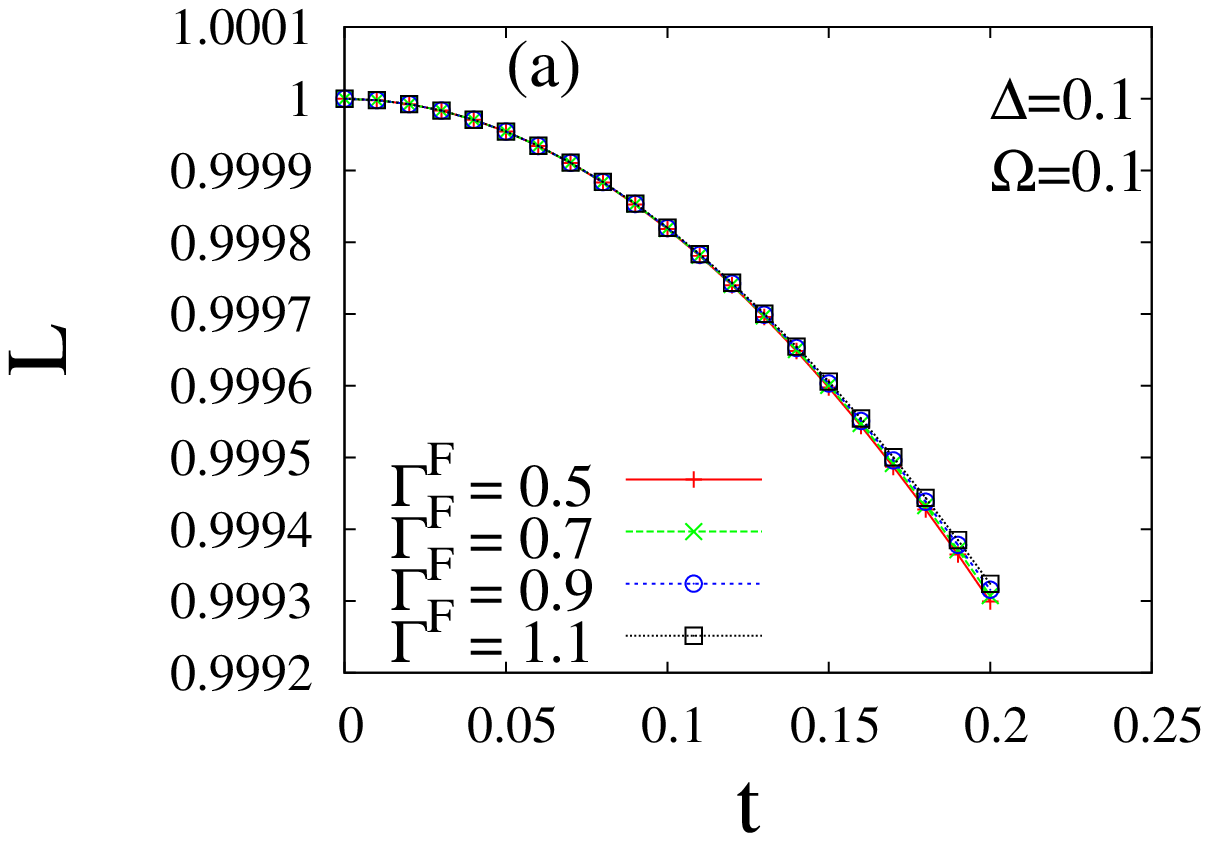}
\includegraphics[width=5.5cm]{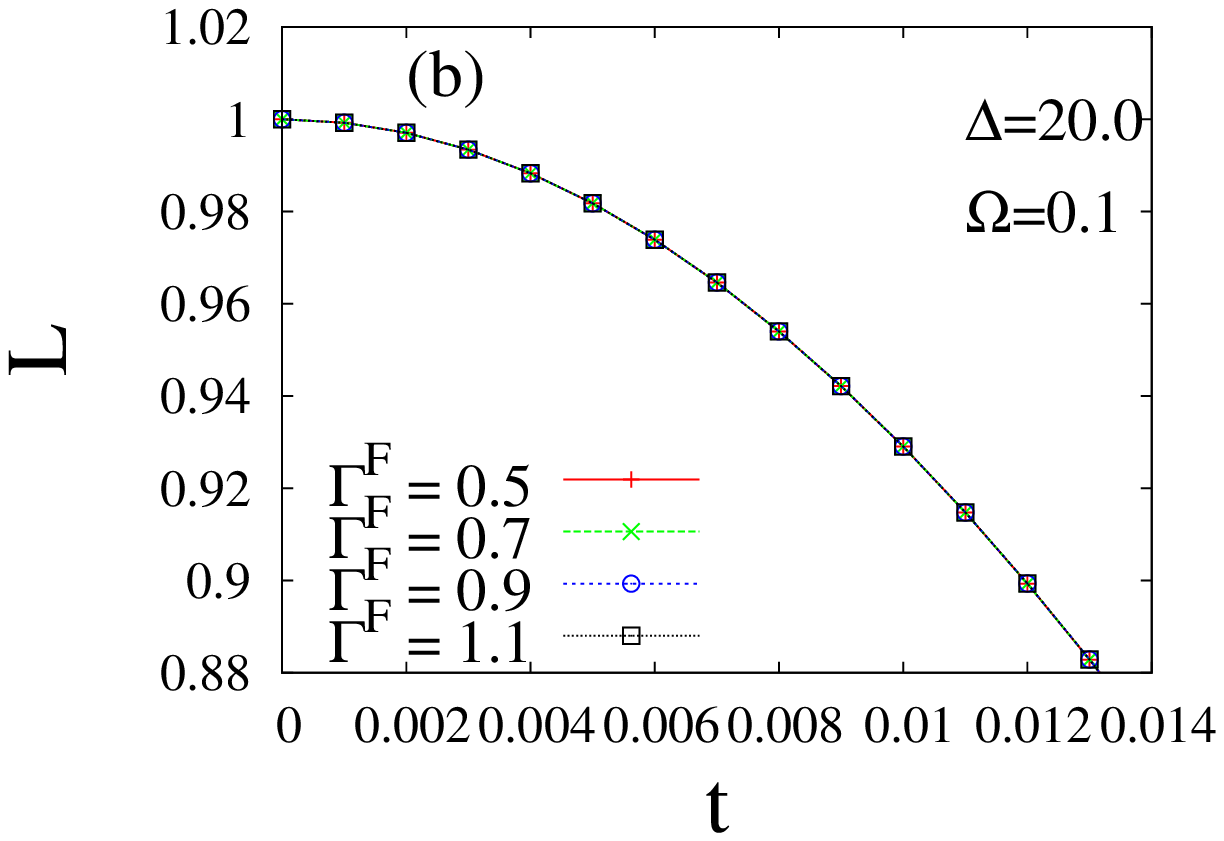}
\end{center}
\caption{(Color online) The ultra short time evolution of Loschmidt echo $L$ for different quenching schedules with coupling 
strength $\Delta = 0.10$ are shown in (a). The initial value of the transverse field is $\Gamma^I = 0.70$ and the final quenched 
values are $\Gamma^F = 0.50, 070, 0.90, 1.10$. The ultra short time dynamics of $L$ for $\Delta  = 20.0$ is shown in (b) with the 
same set of quenching parameters as discussed in (a). In both the cases, the disorder strength is $\Omega = 0.10$.}
\label{low-high-delta-short}
\end{figure*}

First, we start by investigating the behavior of $L$ when the spin chain is quenched from $\Gamma^I=10.0$ to $\Gamma^F=2.0$ 
i.e., both the initial and the final values of transverse fields belong to the paramagnetic phase. The time evolution of $L$ 
for different values of $\Omega$ are depicted in the Fig.~(\ref{para_para}a) and Fig.~(\ref{para_para}b) with weak coupling 
strength $\Delta = 0.1$ and strong coupling strength $\Delta = 10$, respectively. For lower values of coupling strength, we 
find two different types of behaviors of $L$ in two different time domains (see Fig.~(\ref{para_para}a)). Initially echo for 
disordered chain falls of more rapidly as compared to the clean chain; the fall becomes more sharper as one increases the 
disorder strengths. On the other hand, in the late time limit, echo for the clean chain vanishes while for finite $\Omega$ it 
remains non-zero; surprisingly, the  long time value of $L$ increases with $\Omega$.  For strong coupling case, $L$ decays more 
rapidly for less disordered spin chain in both the early and late time limit (see Fig.~(\ref{para_para}b)).

The initial fall is mainly determined by the difference in the initial and final Hamiltonian. In the present set up of GCSM, 
the weak coupling can only modify the transverse field in  two sites in addition to the global sudden quench. Therefore, as 
one increases $\Omega$ from $0$, $H_{\upuparrows}$ and $H_{\downdownarrows}$ become more deviated from each other. Hence, 
the overlap between the initial state i.e., eigenstate of $H_{\downdownarrows}$ and the time evolved state governed by 
Hamiltonian $H_{\upuparrows}$ rapidly decays from unity. In contrary, during the course of late time dynamics the instantaneous 
state evolves more close to the initial state when $\Omega$ increases; the clean chain does not lead to an instantaneous time 
evolved state that significantly close to the initial state. We note that the initial sharp fall is exponential while late time 
slow fall obeys a power law. Therefore, the advantage of having disorder in transverse field is that the qubits remain in an 
entangled state even long after the sudden quench. Moreover, the local details of the two Hamiltonians, governing the dynamics, 
are maximally determined by $\Delta$ for strong coupling case; this results in a similar kind of fall characteristics for both 
early and late time limit.  Disorder imprints less effects on LE for strong coupling case compared to weak coupling case.

Upon inspecting the weak and strong coupling case simultaneously, one can say that the characteristic nature of decay of LE is 
different for them. Although, in presence of disorder, one can see that after the initial quick fall of $L$ it decreases very 
slowly with time; this feature is universal for weak as well as strong coupling limit. Therefore, it is clearly evident after 
investigating the time evolution of LE in complete  time domain that there is a crossover present in the temporal behavior.

\begin{figure*}[htb]
\begin{center}
\includegraphics[width=6.0cm]{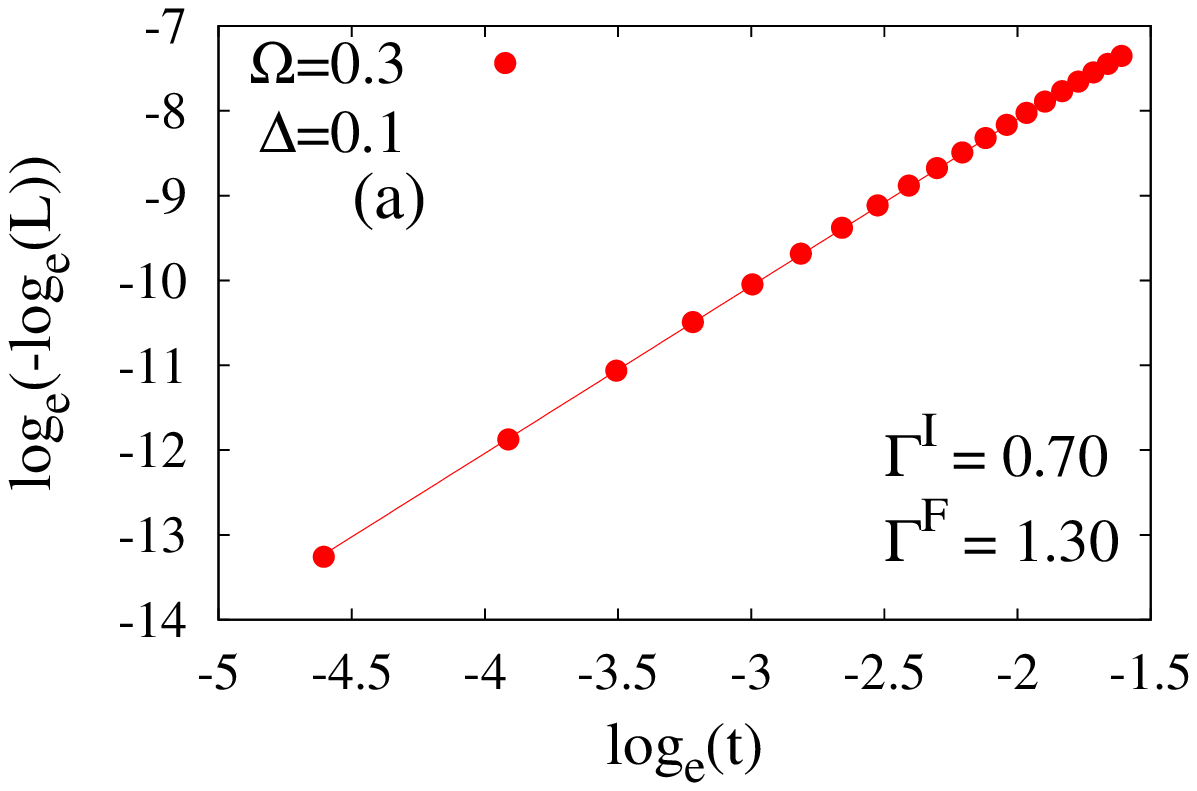}
\includegraphics[width=6.0cm]{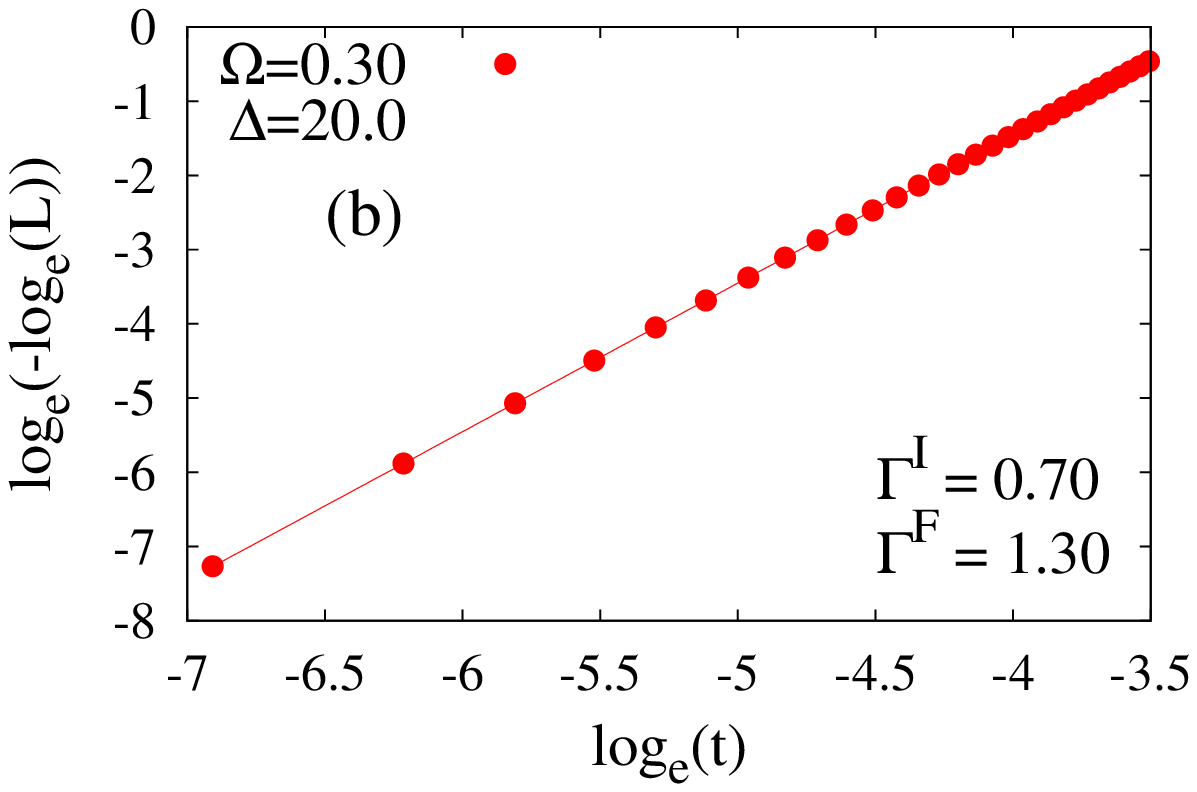}
\includegraphics[width=6.0cm]{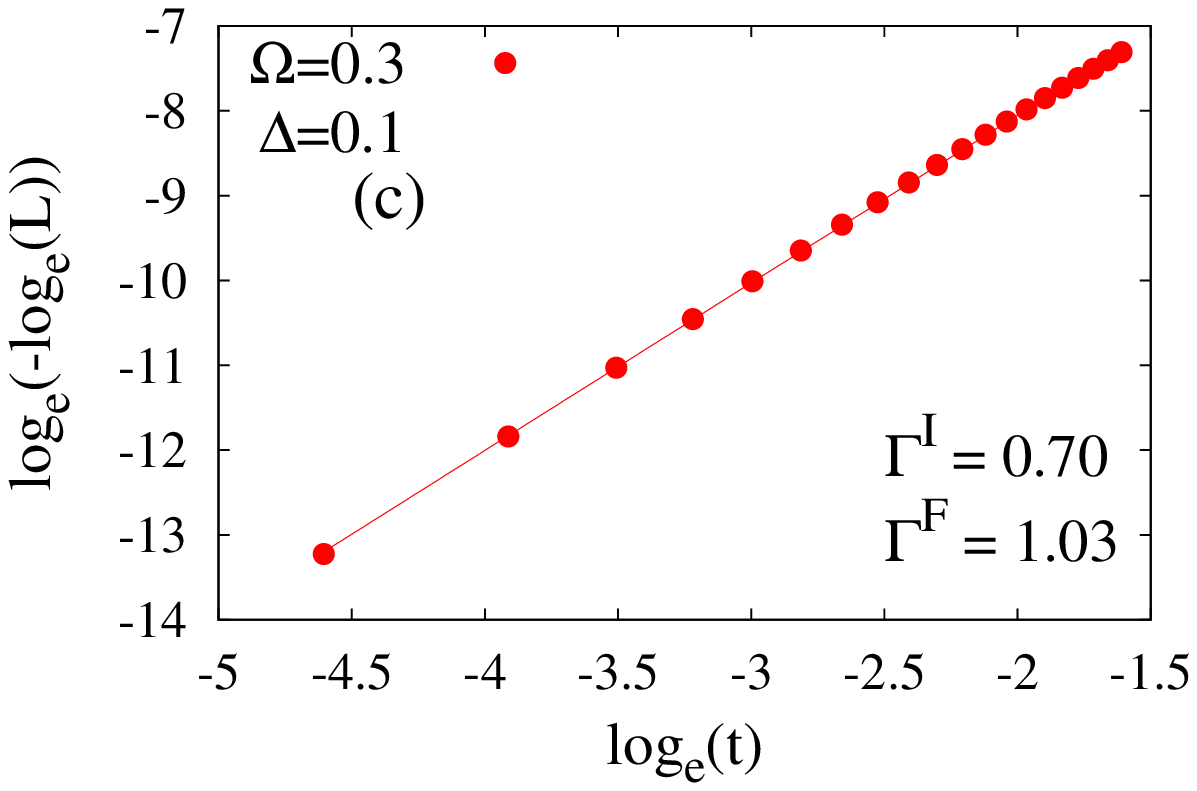}
\includegraphics[width=6.0cm]{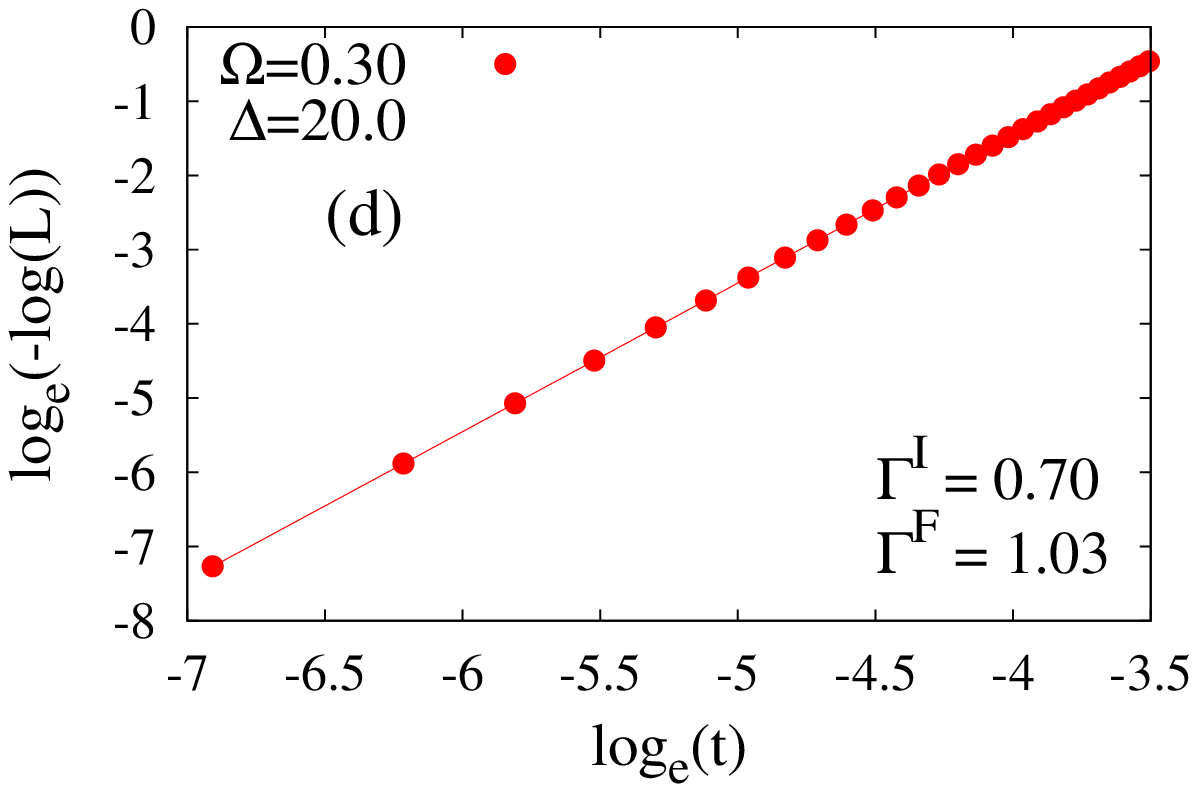}
\end{center}
\caption{(Color online) 
We here depict the ultra short time Gaussian fall of $L$ as observed in (a) for $\Delta=0.1$ and (b) for $\Delta=10.0$ following 
a quench from ferromagnetic phase $\Gamma^I=0.7$ to paramagnetic phase $\Gamma^I=1.3$. The similar ultra short time Gaussian fall 
is observed in (c) for  $\Delta=0.1$ and (d) for $\Delta=10.0$ following a quench from ferromagnetic phase $\Gamma^I=0.7$ to inside 
the Griffith phase $\Gamma^I=1.03$. The slopes of the fitting line, drawn in $\log_e(-\log_e(L))$ vs $\log_e(t)$ plain in all the 
above cases, are closely given by $2$.}
\label{para_pure_gaussian}
\end{figure*}

Now, coming back to the strong coupling case where the time evolution of LE again shows the early time rapid fall and late time 
slow fall (see Fig. (\ref{para_para}b)). The initial decay rate of LE  for early time is higher as compared to weak coupling case. 
This is due to the fact that initial and final Hamiltonian become more deviated for strong coupling case. For clean system after 
the initial fall $L$ shows irregular oscillation which are suppressed in the presence of disorder. The late time dynamics is mainly
governed by disorder strength as the decay nature of LE in small and large $\Delta$ limit are identical.

We shall now turn our attention to the behavior of $L$ when the spin chain is quenched from the paramagnetic phase with $\Gamma^I=10.0$
to ferromagnetic phase  with $\Gamma^F=0.5$ across the Griffiths phase considering $\Delta=0.1$  (see Fig.\ref{para_para}c) and 
$\Delta=10.0$ (see Fig.\ref{para_para}d). In this type of quenching, the initial decay rate is much higher compared to the same 
phase quenching as the eigenstates in the ferromagnetic side become almost orthogonal to the ground states associated with the 
paramagnetic phase. Unlike the quenching inside same phase, here in contrary for both type of coupling, we see that late time value
of LE becomes vanishingly small. The clean system shows zero overlap after a certain time while LE for disordered system still remains 
finite. In case of $\Delta = 0.1$, the values of $L$ for both pure and disordered spin chain falls drastically within a very small 
passage of time. Such time scale is independent of the disorder strength. For large $\Delta$, this time scale is much shorter and 
surprisingly, the clean system also decays almost identically with the disordered system. After the initial rapid fall for strong 
coupling case, there are oscillations in  $L$ for both clean and disordered spin chain in the late time regime with very slow fall; 
the amplitude of oscillation for pure system is larger than the disorder spin chain. Although, such oscillations die very quickly 
for both types of spin chains.

Now, we shall investigate the temporal evolution of LE in different time domains following various quenching amplitudes. Let us 
begin with the ultra-short time analysis of the LE to investigate the fall of $L$ from unity.  We numerically study ultra-short 
time evolution of LE for both low ($\Delta = 0.10$) and high ($\Delta = 20.0$) values of coupling strength with $\Omega = 0.10$. 
For both $\Delta = 0.10$ (see Fig.(\ref{low-high-delta-short}a)) and $\Delta = 20.0$ (see Fig.(\ref{low-high-delta-short}b)) the 
transverse field is quenched from $\Gamma^I = 0.70$ to the final value $\Gamma^F = 0.5, 0.70, 0.90, 1.10$. In this extremely short
time window, LEs for strong and weak coupling following different quench amplitudes coincide with each other.

Our numerical result  shows that in both the situations, a Gaussian fall is observed within a certain  time scale that depends on
$\Delta$. The fitting of $L$ in ultra-short time window for ferromagnetic to paramagnetic quench with $\Delta=0.1$ and $\Delta=20.0$ 
are shown in Fig. (\ref{para_pure_gaussian}a) and Fig. (\ref{para_pure_gaussian}b), respectively. The typical characteristic time scale 
for $\Delta \ll 1$ is given by $\hat t= 1$, whereas, in case of high values of coupling strength i.e., $\Delta \gg 1$, the value of 
characteristic time is $\hat t= 1/\Delta$. The ultra-short time the decay of $L$ follows a Gaussian fall which can be approximated as  
$L(t) \simeq \exp(-\alpha t^2 )\approx 1 - \alpha t^2$. Here, $\alpha$ determines the rate of the fall and the power of $t$ i.e., $2$
signifies the exponent for the  Gaussian decay. We also check that this decay rate $\alpha$ is marginally dependent of $\Omega$ for both 
type of coupling. As expected, the rate $\alpha$ for strong coupling is much higher compared to the weak coupling case. Our study further
indicates that quenching inside Griffiths phase also lead to this ultra-short time Gaussian fall as depicted in Fig. (\ref{para_pure_gaussian}c) 
for $\Delta=0.1$ and Fig. (\ref{para_pure_gaussian}d) for $\Delta=20$. {Such Gaussian decay in ultra-short timescale for clean system is
already observed in \cite{wendenbaum14}.} Therefore, one can say that the Gaussian fall observed in the ultra-short time domain remains
invariant even in presence of disorder.

It has been found for a non spin preserving coupling Hamiltonian $H_Q$ that rate $\alpha$ for the Gaussian decay is dependent on 
the filling number of the ground state of the bath; specifically, $\alpha$ exhibits plateaus as a function of intra bath coupling 
and transverse field \ct{wu14}. Using time dependent perturbation theory it has been further confirmed analytically. The same line 
of argument is valid for our case also only the interaction Hamiltonian changes. Moreover, we find that $\alpha$ is marginally 
dependent on disorder strength and it is evident from fig. (\ref{para_pure_gaussian}) that ultra short time dynamics is mainly 
dictated by the initial parameter i.e., the value of transverse field. Hence, in general, one can say that $\alpha$ strongly depends
on the initial state rather than the final state. To be precise, the rate is shown to be in the form of 
$(\langle H_Q^2 \rangle- \langle H_Q \rangle^2)$, where $H_Q$ is the interaction Hamiltonian between qubit and the bath, $\langle \cdots \rangle$ represents the expectation value with respect to the initial state of the bath \ct{peres84}.

\begin{figure*}[htb]
\begin{center}
\includegraphics[width=5.5cm]{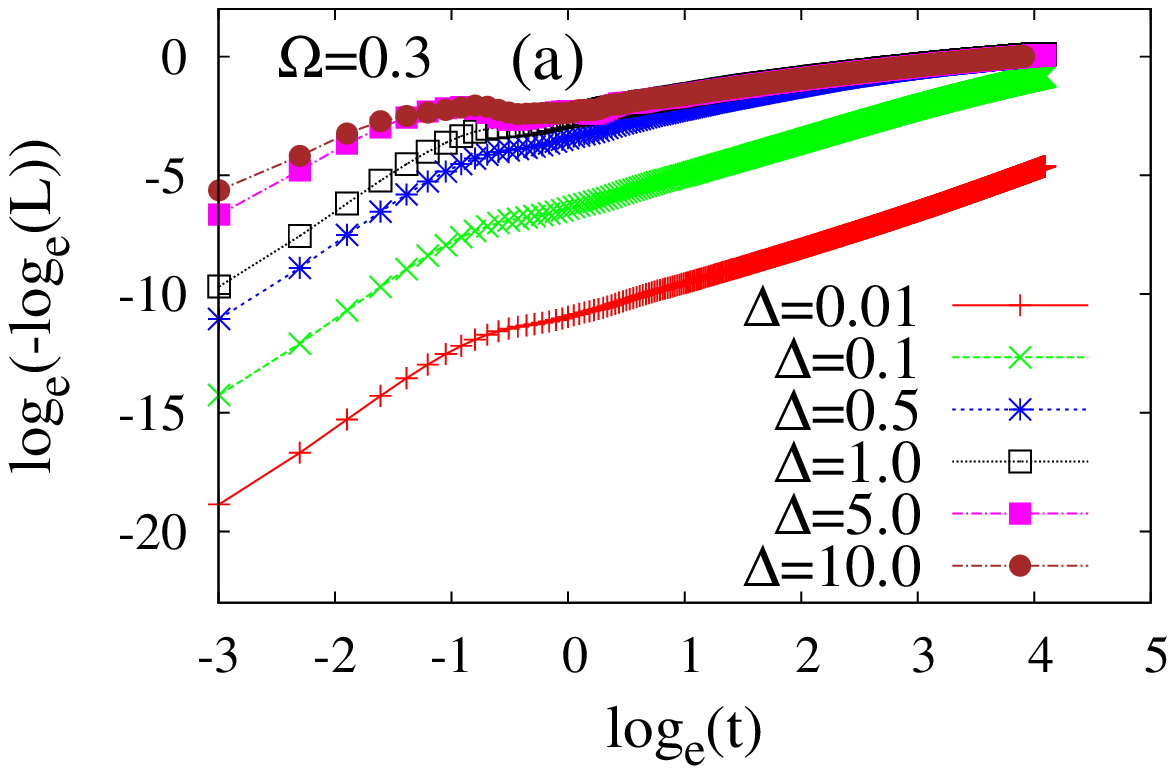}
\includegraphics[width=5.5cm]{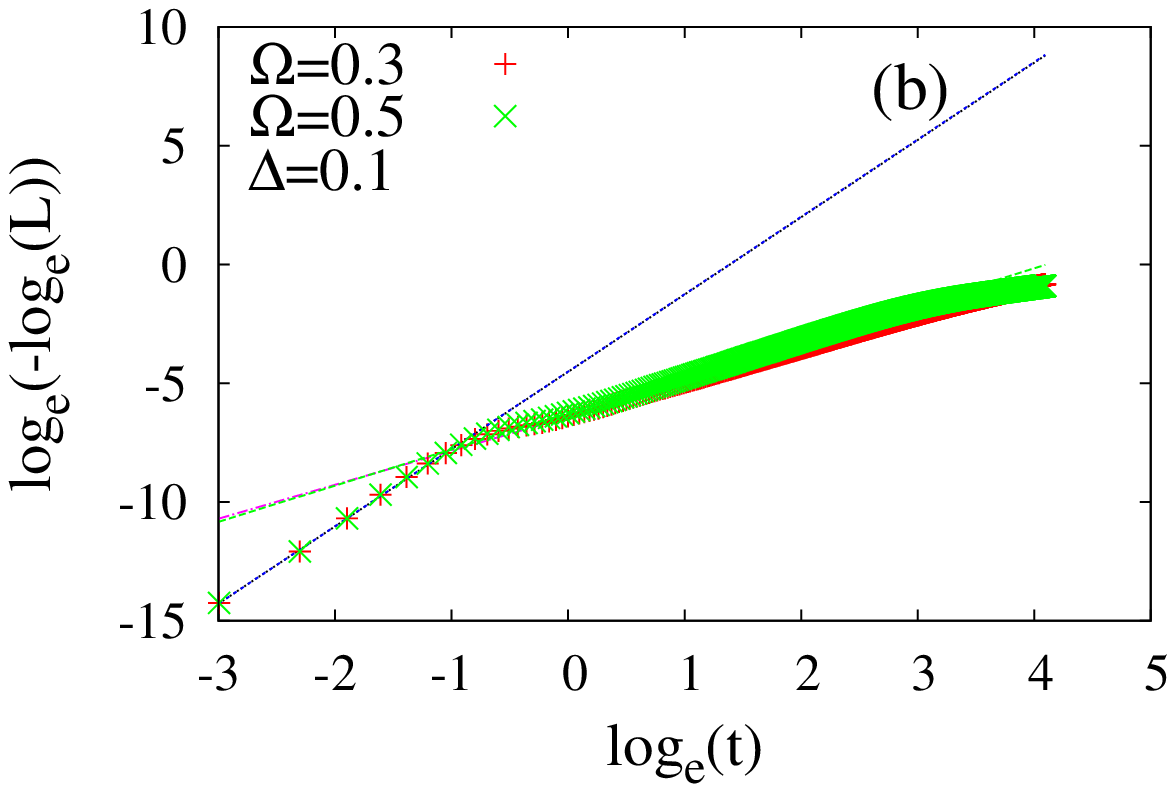}
\includegraphics[width=5.5cm]{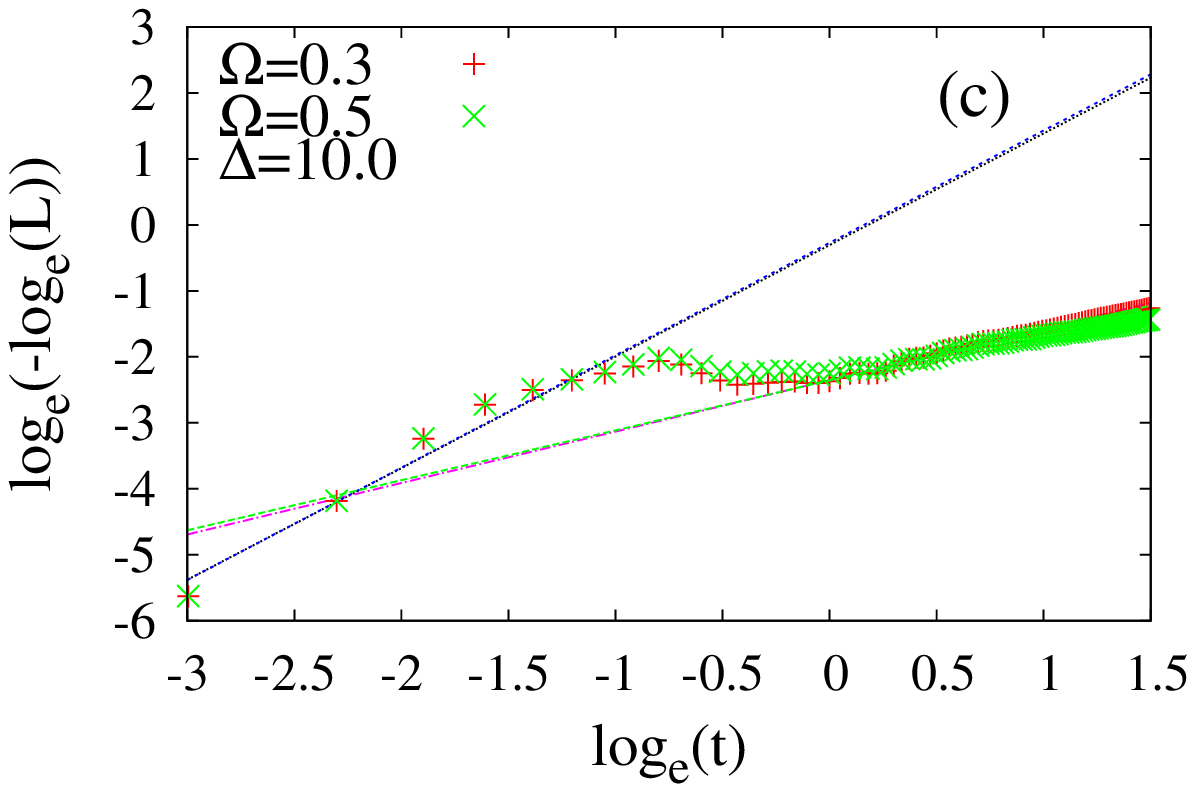}
\end{center}
\caption{(Color online) (a) The plot shows that LE $L$ follows two types of exponential fall with time $t$ for coupling strengths
varying from $\Delta = 0.01$ to $10.0$. These two exponential fall are evident from two different slopes  associated with the initial 
time straight line ($-3<\log_e (t)<-1$) and late time straight line ($1<\log_e (t)<4$), observed in double logarithmic of echo versus
logarithmic of time. It is noteworthy that these slopes acquire two distinct values for $\Delta< 2$ and $\Delta>2$. (b) Plot shows that 
for weak coupling limit ($\Delta=0.1<2$), these two exponential laws in two time domains indeed become independent of $\Omega$. The 
estimated values of the slopes in these time regions are $m \approx 3.25 \pm 0.05 $ and $m \approx 1.50 \pm 0.02$ respectively. (c) 
Plot depicts the exponential behavior quantitatively changes for strong coupling case ($\Delta=10 >2$). The slopes are found to be 
$m \approx 2.00 \pm 0.05$ and $m \approx 0.75 \pm 0.02$. Here the transverse field is quenched from $\Gamma^I=10.0$ to $\Gamma^F=2.0$.}
\label{para_para_all_short}
\end{figure*}

\begin{figure*}[htb]
\begin{center}
\includegraphics[width=5.5cm]{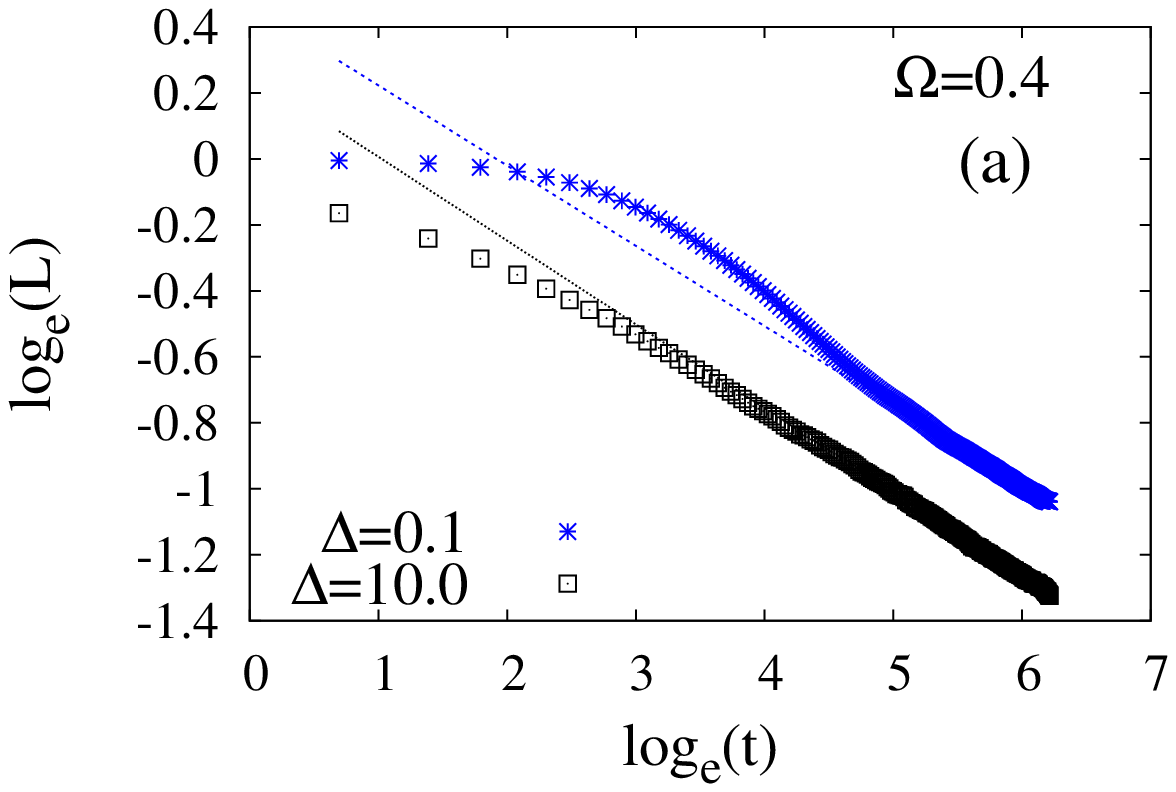}
\includegraphics[width=5.5cm]{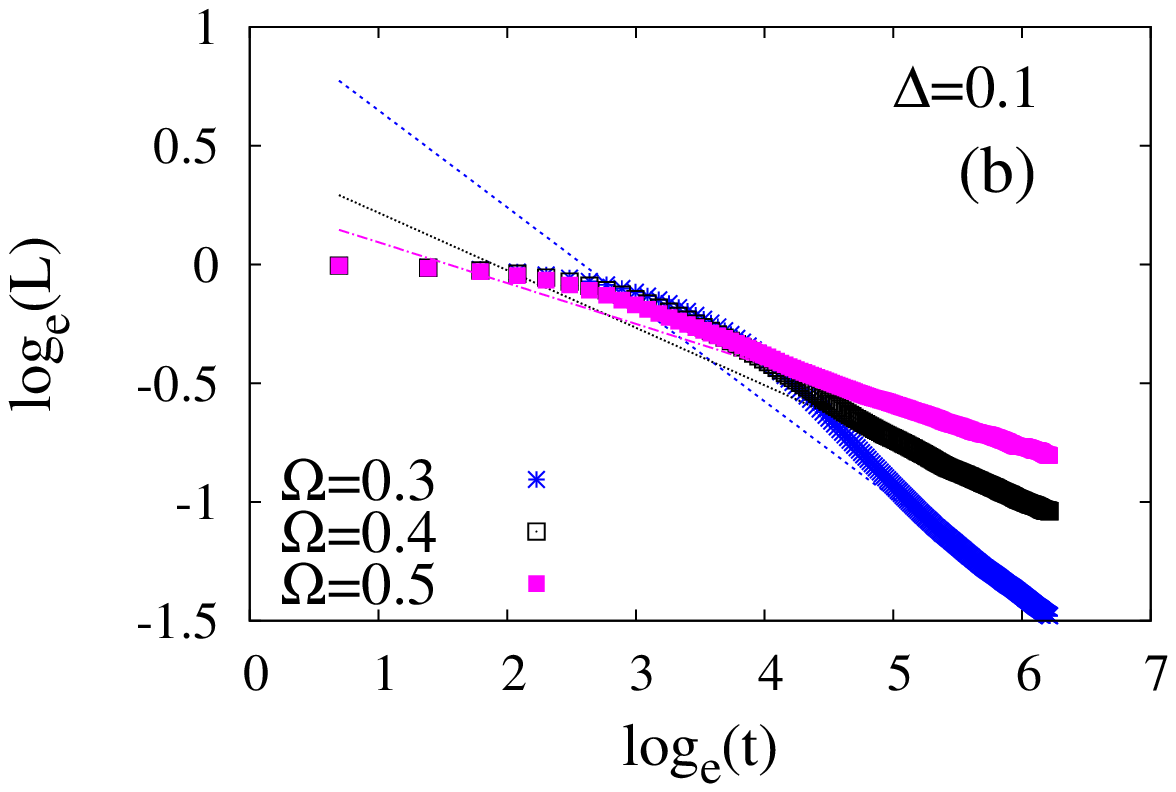}
\includegraphics[width=5.5cm]{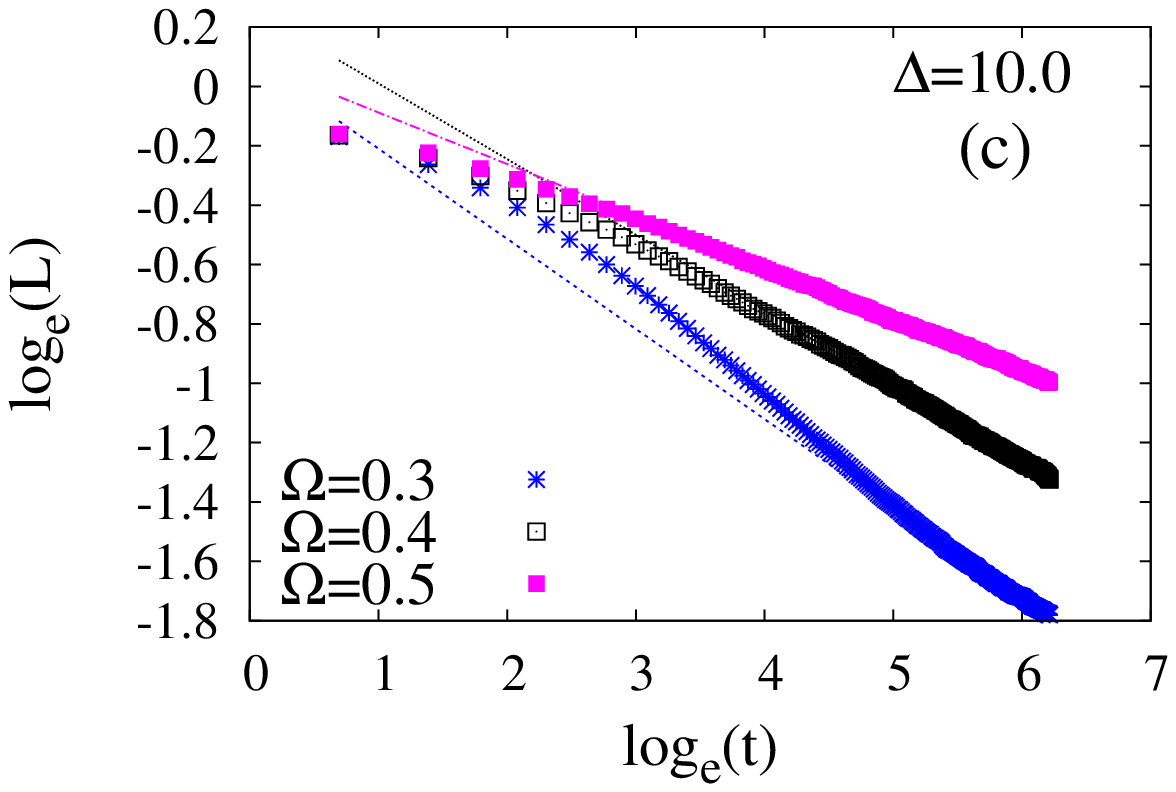}
\end{center}
\caption{(Color online) (a) Plot shows that the LE clearly exhibits long time power law behavior $L(t)\sim c t^{-\theta}$. It is clearly
evident from the parallel nature of the logarithm of echo with different values of $\Delta$ that the power law exponent $\theta$ is 
independent of $\Delta$ rather only depends on $\Omega$. The value of power law exponent for this disorder strength is $\theta \approx 0.25$. 
(b) The variation of $\log_e(L)$ with $\log_e(t)$ for $\Omega = 0.3, 0.4, 0.5$ are shown. Here the coupling strength $\Delta = 0.10$. It
is observed that the magnitude of $\theta$ decreases with the increase of $\Omega$. The values of the exponents are $0.30, 0.24, 0.17$ for
$\Omega = 0.3, 0.4, 0.5$ respectively. (c) The same variations are plotted with identical set of disorder strength. In this case $\Delta = 10.0$. 
We note that $\Omega$ dependence of the power law exponent is quite prominent. In all  of the cases, transverse field  is quenched from
$\Gamma^I=10.0$ to $\Gamma^F=2.0$.}
\label{para_para_all_long}
\end{figure*}

\begin{figure*}[htb]
\begin{center}
\includegraphics[width=5.5cm]{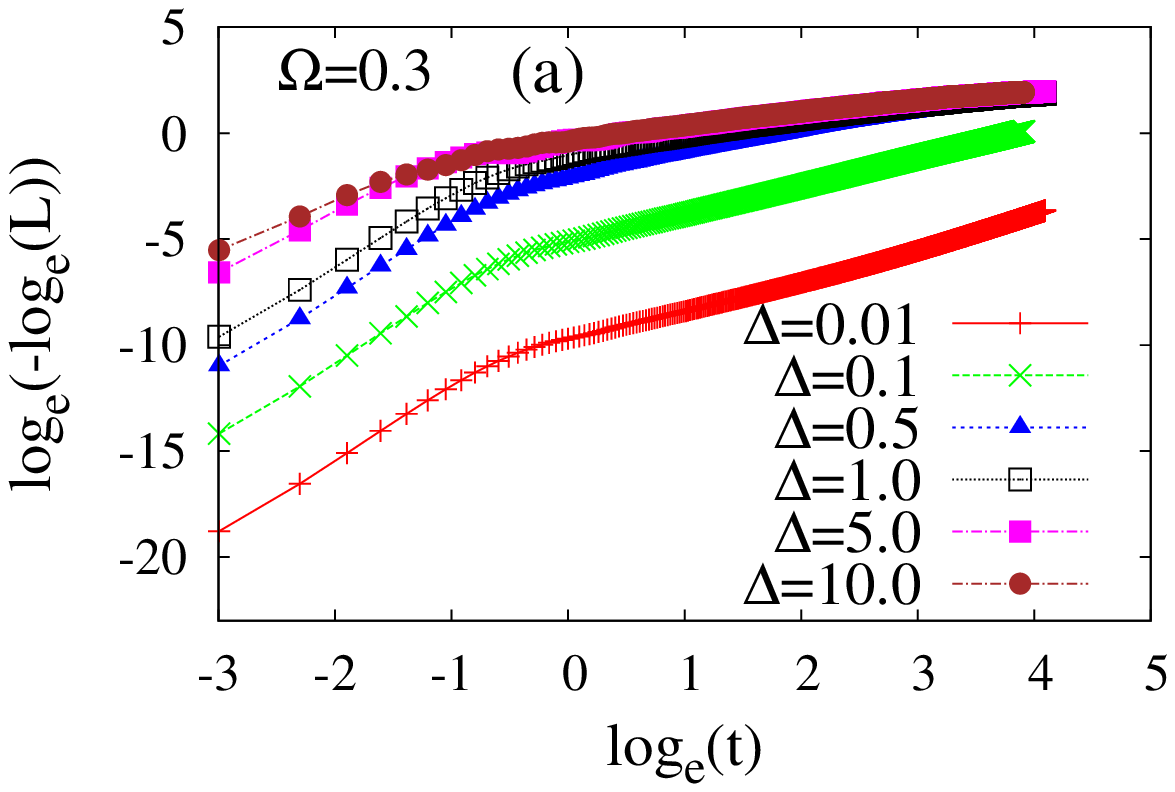}
\includegraphics[width=5.5cm]{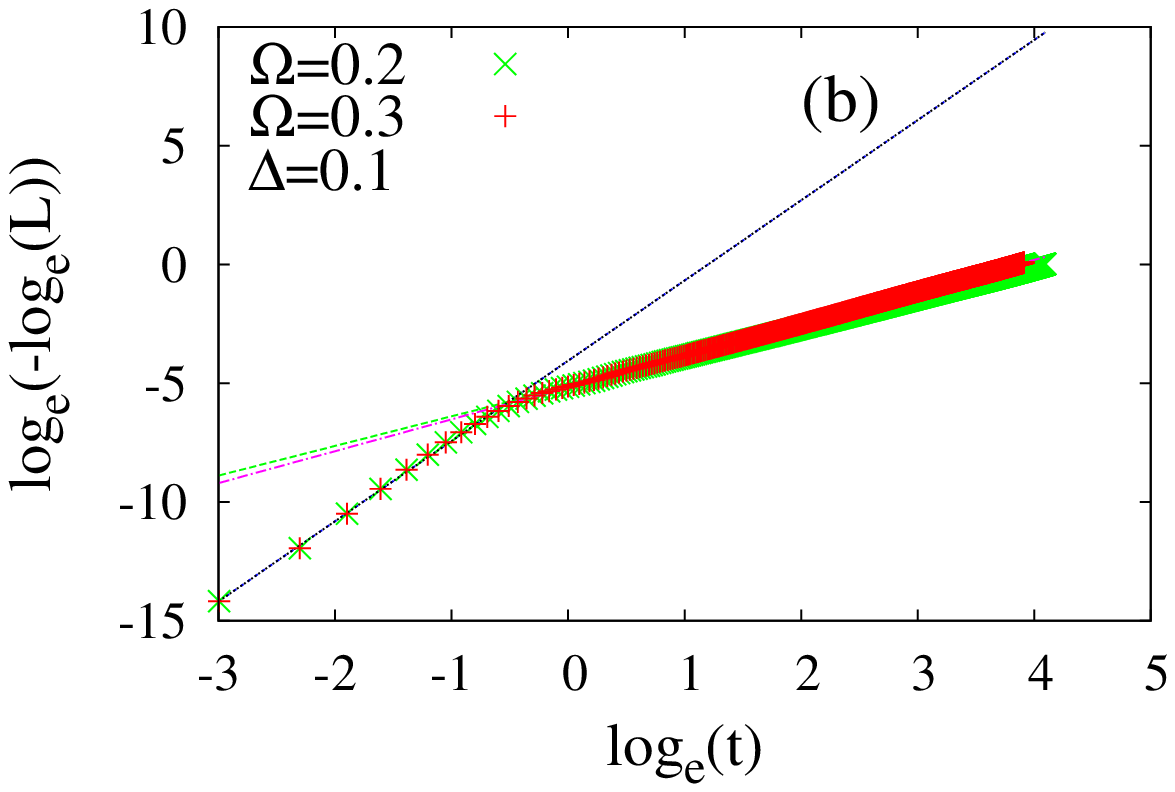}
\includegraphics[width=5.5cm]{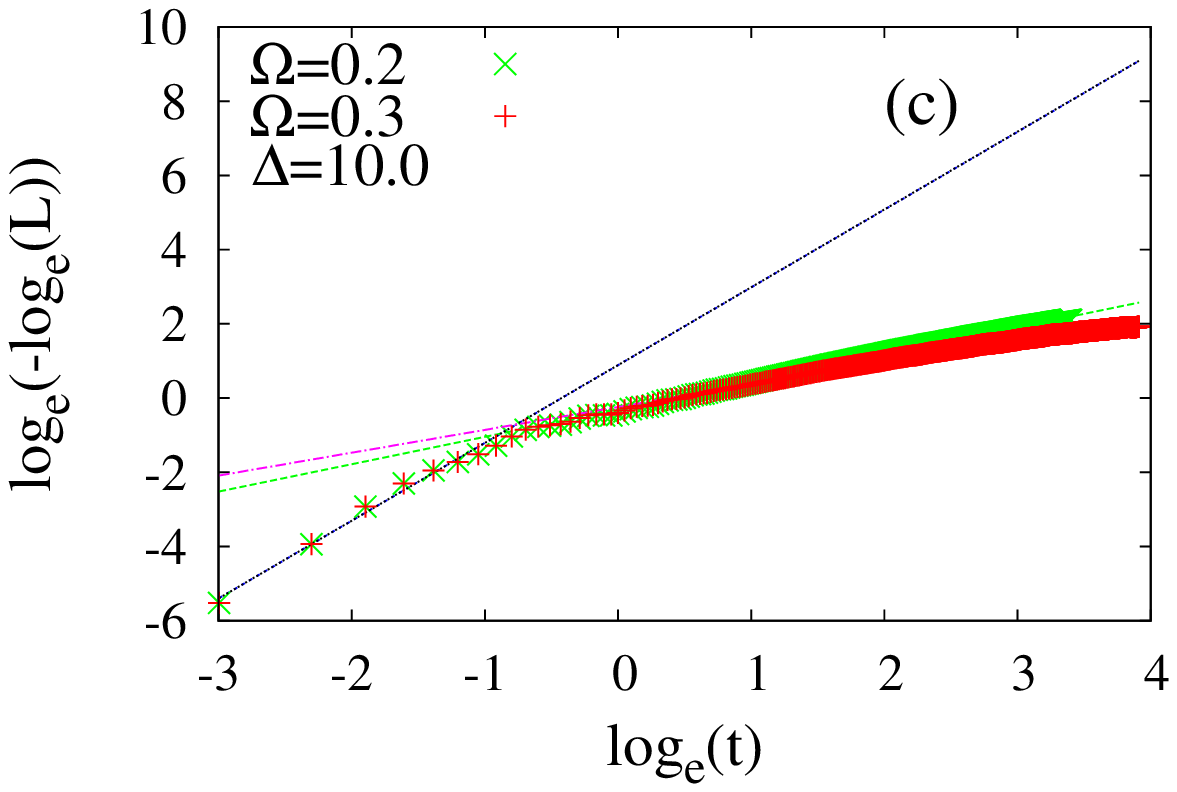}
\end{center}
\caption{(Color online) We repeat the plots as shown in Fig.~(\ref{para_para_all_short}) for the quenching from paramagnetic to 
Griffiths phase. (a) Plot depicts that with a given disorder strength $\Omega = 0.3$, there are two different types of exponential 
fall in the time domains  $-3<\log_e(t)<-1$ and $1<\log_e(t)<4$, respectively. Here, coupling strengths  are varied from $\Delta = 0.01$ 
to $10.0$. (b) Plot shows that for weak coupling limit ($\Delta=0.1<2$),  the slopes indeed become independent of $\Omega$ like the
paramagnetic quenching case. In this case, the values of the exponents associated with the first and second exponential decay are 
$m \approx 3.35 \pm 0.07$ and $m \approx 1.30 \pm 0.02$, respectively. (c) Plot depicts the exponential behavior quantitatively 
changes for strong coupling case ($\Delta=10 >2$). The estimated values of slopes are $m \approx 2.00 \pm 0.05$ and $m \approx 0.70 \pm 0.02$. 
Here, the transverse field is quenched from $\Gamma^I=10.0$ to $\Gamma^F=1.03$.}
\label{para_grf_short_time}
\end{figure*}

\begin{figure}[htb]
\begin{center}
\includegraphics[width=5.5cm]{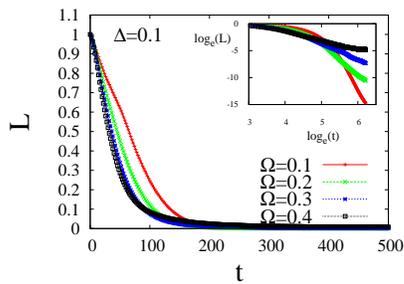}
\end{center}
\caption{(Color online) 
We here show that LE does not exhibit any  slowly falling tail  in the long time limit. The absence of long time power law decay is 
a signature of Griffiths phase. The LE  for weak disorder case decays rapidly as compared to the strong disorder case.The absence of 
power law is further confirmed in the  inset where it is  shown that in logarithm of LE does not follow a straight line with logarithm 
of time. Here the transverse field is quenched from $\Gamma^I=10.0$ to $\Gamma^F=1.03$.}
\label{para_grf_long_time}
\end{figure}

At the outset, we expect that in general the decay exponent (i.e., power of $t$) depends on the disorder strength and the coupling
strength. Upon inspecting the Fig.~(\ref{para_para_all_short}) and Fig.~(\ref{para_para_all_long}), one can remarkably see for 
quenching inside the same phase that for the initial small time exponential fall, the decay exponent depends only on the coupling 
strength; on the other hand, the decay exponent associated with the  late time power law fall depends only on the disorder strength. 
The quenching inside a Griffiths phase as shown in Fig.~(\ref{para_grf_short_time}) and Fig.~(\ref{para_grf_long_time}) leads to a 
markedly different results in the long time regime; the power law fall is completely absent here.

In order to investigate the temporal decay of LE with  disorder for the quenching inside paramagnetic phase ($\Gamma^I=10.0$ to $\Gamma^F=2.0$)
more extensively, we first study the initial small time rapid  decay. In such time scale the double logarithm of LE becomes linear
with the logarithm of time. This indicates the small time faster fall is exponential in nature. Our numerical analysis further 
suggests that there exist more than a single exponential fall; LE scales as $L(t)\sim \exp(-\beta t^m)$ and the decay exponent $m$ 
takes two different values in two different time regions confined within this overall exponential time zone. The decay rate is encoded 
in $\beta$. The values of $m$ in such time domains depend on $\Delta$ mainly. To reveal such feature, we plot $\log_e(-\log_e(L))$ with 
$\log_e(t)$ for a fixed disorder strength $\Omega = 0.10$ with several values of $\Delta$ (see Fig.\ref{para_para_all_short}a). Inside
each characteristic region of time, two sets of parallel lines are obtained for $\Delta < 2$ and $\Delta > 2$. For weak coupling, $m$ 
changes from  $m \sim 3.25$ to  $m \sim 1.50$ (see Fig.\ref{para_para_all_short}b). In the strong coupling limit, LE continues to follow 
a Gaussian fall i.e., $m \sim 2.0$ and in later time the decay exponent is close to $0.75$ (see Fig.\ref{para_para_all_short}c). One can
clearly observe the LE for different $\Omega$ (for a given $\Delta$) almost co-inside with each other. This refers to the fact that decay
exponent does not change with the disorder strength.

We shall now focus on the late time behavior of LE for the quenching inside the paramagnetic phase. We quench the transverse filed 
$\Gamma^I=10.0$ to $\Gamma^F=2.0$. We observe a late time slow power law fall of LE. Such fall can be described by $L(t)\sim c t^{-\theta}$, 
where $c$ determines the decay rate, and $\theta$ is the decay exponent for the power law. This late time power law decay of LE is 
universal for weak and strong coupling case but time domain within which this power law observed is dictated by the value of the 
coupling strength. To demonstrate such characteristic,  we plot $\log_e(L)$ with $\log_e(t)$ for $\Delta = 0.1, 10.0$ with fixed 
disorder strength $\Omega = 0.4$ (see Fig.\ref{para_para_all_long}a). One can clearly find that the straight lines corresponding to 
different values of $\Delta$ become parallel to each other; this suggests the  power law exponent is indeed independent of coupling 
strength. In order to investigate  $\Omega$ dependence of $\theta$, we plot the variation of $\log_e(L)$ with $\log_e(t)$ for several 
values $\Omega$ keeping  coupling strength fixed at $\Delta = 0.1$ (see Fig.\ref{para_para_all_long}b) and $\Delta = 10.0$ 
(see Fig.\ref{para_para_all_long}c). Irrespective of the value of coupling strength, we find that magnitude of $\theta$ increases 
with the decrease of $\Omega$. Moreover, we checked that the decay characteristics of LE (as described by exponential and power law 
fall) for quenching into a ferromagnetic phase from the paramagnetic phase remain unaltered. However, we observed a change in the 
extent of the time domains associated with the different types  of decay.

{We shall now try to justify the above numerical finding by plausible arguments. The $\Delta$ dependent initial exponential 
fall is consistent with the analytical results in Refs.~\ct{sudip_Rossini2007,wendenbaum14}. The analytical treatments suggest the 
decay rate is of the order of ${\Delta}^2$. Therefore one would expect the initial exponential fall is dependent on coupling strength 
and the initial value of the transverse field rather than the disorder strength.  On the other hand the asymptomatic behavior of LE 
depends on the final value of the quenched transverse field~\ct{sudip_Rossini2007}.  In a given quenching scheme, the disorder strength 
determines the final quenched values of transverse field at each spin site. Therefore the late time fall of LE should be governed by 
the disorder strength. Moreover, for strong disorder, the difference between the  initial and final values of the local transverse 
field becomes small. As a result of that the decay of LE is severely prohibited with the increment in disorder strength. Such phenomena 
is exactly reflected by our numerical study associated to the late time fall of LE.} 

In parallel, we study the quenching inside the Griffiths phase starting from a paramagnetic phase. Similar to the paramagnetic quenching
case, the initial time evolution of LE is governed by two types of exponential fall: $L(t)\sim \exp(-\beta t^m)$ with two different values 
of $m$. The decay exponent again predominantly depends on $\Delta$ rather than $\Omega$ (see Fig.\ref{para_grf_short_time}a). In 
case of weak coupling, the LE falls off initially with $m \sim 3.35$ and the subsequent exponential decay is accompanied with $m \sim 1.30$ 
(see Fig.\ref{para_grf_short_time}b). For strong coupling case, the primary exponential fall is Gaussian with $m \sim 2.0$ and the final 
exponential decay is associated with $m \sim 0.70$ (see Fig.\ref{para_grf_short_time}c).

Most interestingly, we find that LE decays to zero in the late time limit (see Fig.~(\ref{para_grf_long_time})). This result can
be contrasted to the earlier results obtained for off-critical quenching between same or different phases. In order to confirm 
the absence of the power law tail, logarithm of LE is plotted  as a function of logarithm of time; this plot clearly depicts that
logarithm of LE does not fit into a  straight line within a considerable time domain. The point to note here is that as the slowly
falling power tail is absent the LE decays to zero more rapidly compared to the other quenching cases. Therefore, the absence of the
late time power law fall is a dynamical signature of the  Griffiths phase. One can see that a tendency towards a late time slow
power law fall is observed for higher values of $\Omega$ as for some disorder realizations, environmental Ising chain lies outside 
the Griffiths phase.

The outcome quantum Griffiths effect is that the rare region is finite in space but infinite in imaginary time~\ct{vojta14}; 
moreover, the broadening in the distribution of local relaxation time is another signature of Griffiths singularity. 
Consequently, the effect of the local perturbation (i.e., coupling to the qubit locally) is damped inside this huge temporal window. 
This might cause the initial exponential fall to prevail over the late time power law fall. The quantum Griffiths effect is absent 
in the PM and FM phase. The distribution of local relaxation time gets narrower leading to the fact that the effect of local perturbation
is not damped completely rather it is still present in the long time. This might leads to the observation of long time power law fall
of LE. This is the way we think that the characteristics of the spin chain gets imprinted the behavior of LE. Therefore,  concurrence between 
the qubits is  thus able to distinguish the phase of environmental spin chain associated with a continuously varying dynamical exponent.

Comparing the short time  temporal behavior (i.e., exponential fall) of LE for different quenching case one can see that when the 
initial and final phases are different, LE drops most rapidly (see Fig.\ref{para_para}c,d), while LE drops most slowly for same initial 
and final phases (see Fig.\ref{para_para}a,b);  for quenching into Griffiths phase from paramagnetic phase, LE attains an intermediate 
fall (see Fig.\ref{para_grf_long_time}). Although, the functional form of decay remains exponential, the decay rate, estimated from
$\beta$ in $L(t)\sim \exp(-\beta t^m)$, generically depends on the disorder strength, types of quenching and coupling strength. This 
decay rate is highest for quenching between two different phases and smallest for quenching between same phases. Moreover, this decay 
rate increases with increasing disorder strength in the weak coupling case. The same line of argument is also valid for decay rate 
associated with late time power law fall. We press the fact that disorder maximally influences the late time slow fall than the short 
time exponential fall, dominated by $\Delta$, as for the clean spin chain, LE decays to a much lower value in this late time limit. 
Hence, disorder imprints a positive effect in preserving an entangled state surrounded by a decohering environment.

\begin{figure*}
\begin{center}
\includegraphics[width=5.5cm]{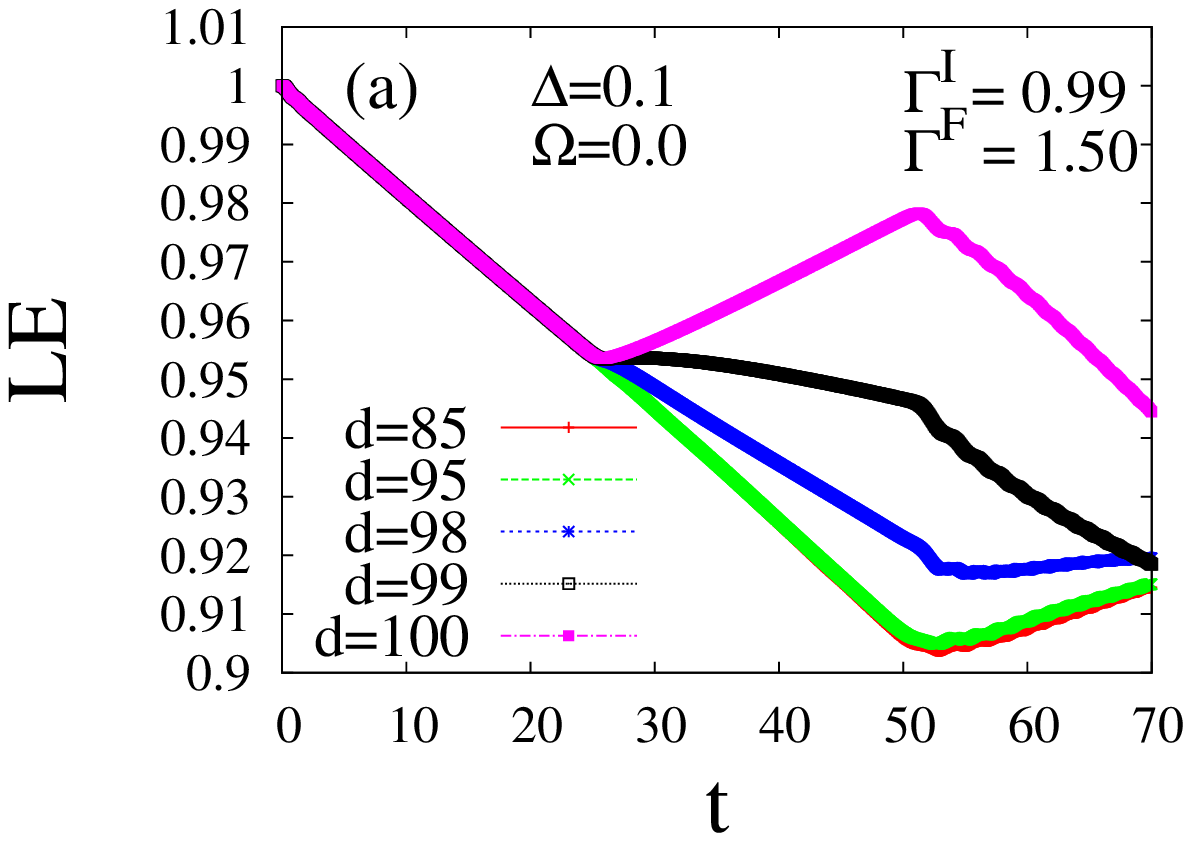}
\includegraphics[width=5.5cm]{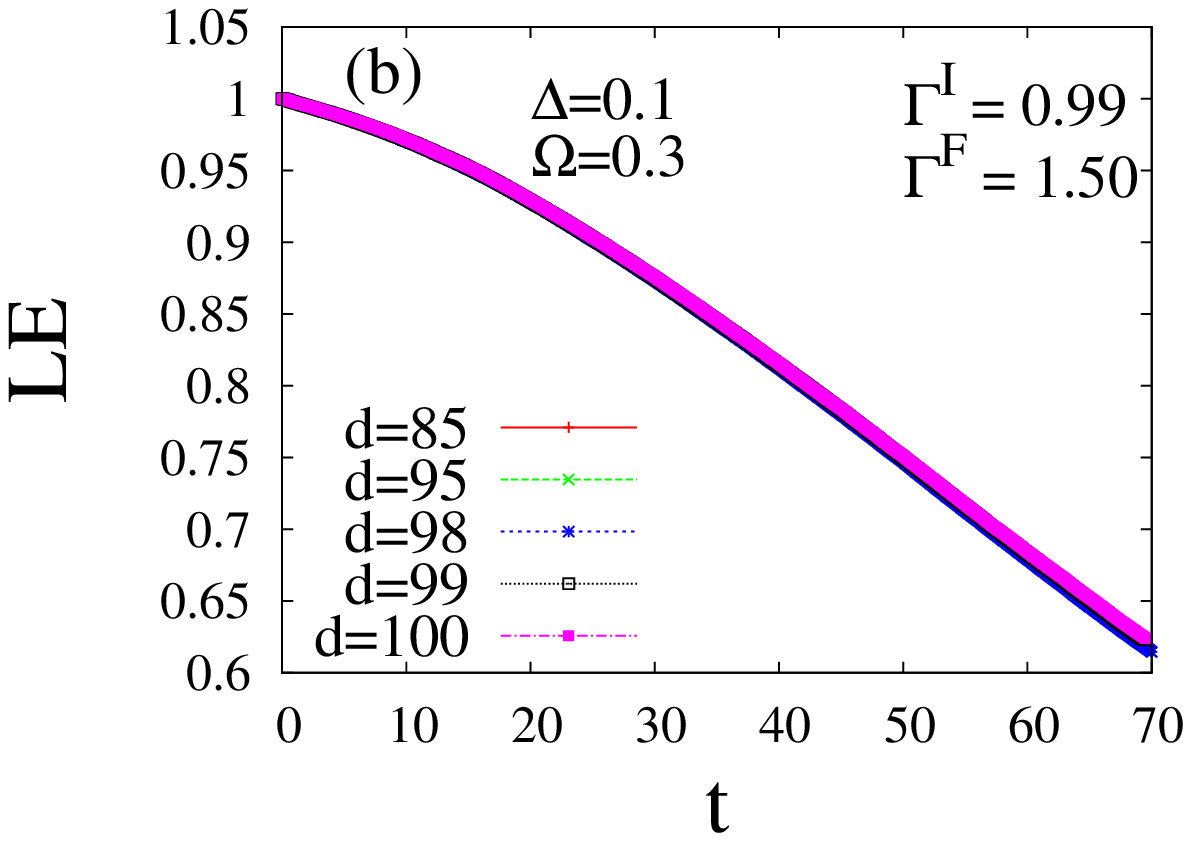}
\includegraphics[width=5.5cm]{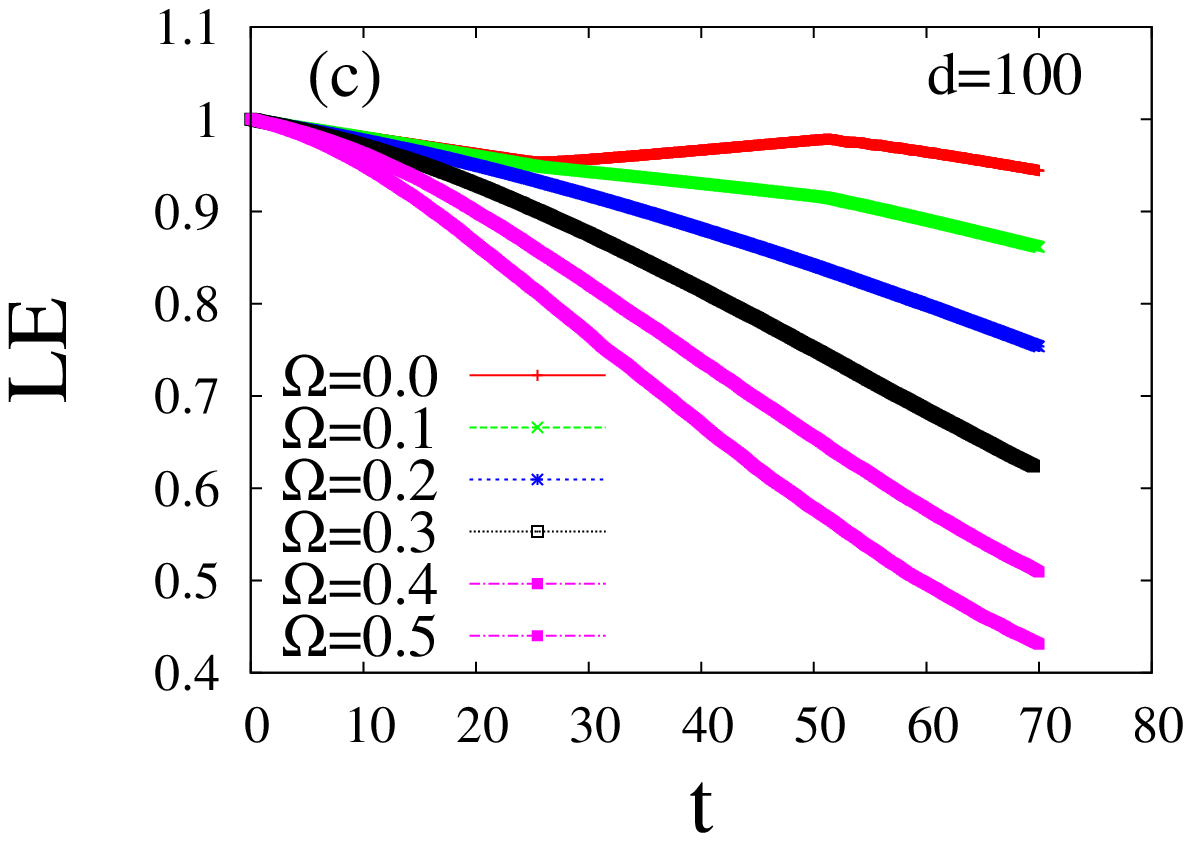}
\end{center}
\caption{(Color online) The variation of $L$ with time $t$ for a given strength of disorder are shown: (a) for $\Omega = 0.0$ 
(b) for $0.30$ with different values of separation distance of two qubits $d = 100, 99, 98, 95, 85$. (c): The variation of $L$ 
with time $t$ for different values of $\Omega = 0.0, 0.10, 0.20, 0.30, 0.40, 0.50$ are shown. Here we fixed the separation distance
$d = 100$. We can see for $\Omega > 0.10$ there are no singularity in the slope of $L$. We quenched the transverse field 
$\Gamma^I = 1.50$ to $\Gamma^F = 0.99$.}
\label{revival_time}
\end{figure*}

\subsection{Revival Time}
In addition to the coupling strength, the dynamics of $L$ is also governed by the separation $d$ between two qubits coupled 
to the environmental spin chain. In this section, we discuss the dynamics of $L$ for both clean and disorder spin chain of 
length $N = 200$. Before going to the disordered chain, we shall quickly review the clean case for different values of $d$
following a quench from $\Gamma^I = 1.50$ to $\Gamma^F = 0.99$ as shown in Fig.~(\ref{revival_time}a). {We should 
mention that such investigation has already been made with $N = 100$ in Ref~\ct{wendenbaum14} but we repeat (with different system size) 
this numerical calculation to present a comparative study between the observations associated to the revival of LE in pure and disordered
spin chain.} When the qubits are connected symmetrically i.e., 
$d = N/2$,  we find that echo exhibits a linear decrease till $t=\tau_1 \simeq N/8$; afterwards, of $L$ revives linearly till 
$t=\tau_2 \simeq N/4$. As one moves away from the symmetric position by changing the separation distance from 
$d = N/2$ to $d = N/2 -15$, the time scale $\tau_1$ gradually disappears.

Concentrating on the disordered case, we show the variation of $L$ with $t$ for $\Omega = 0.30$ with several values of $d$ 
as depicted in Fig.~(\ref{revival_time}b). One can observe that a substantial amount of disorder can wash out the above 
dynamical characteristics  appearing in  $L$ at $t=\tau_1$ {and  $t=\tau_2$}. The revival of LE after $t=\tau_1$ and the 
downturn {at  $t=\tau_2$} as obtained in pure case are absent in the disordered spin chain. In this case, LE monotonically 
decreases with time. The LE for symmetric case exhibits a higher value as compared to the non-symmetric position. 
In order to investigate the disappearance of singular behavior of LE, we study the LE for different disorder strength with 
{a} given $d=N/2$. This is pictorially shown in Fig.~(\ref{revival_time}c). One can see that as $\Omega$ increases, the revival 
of LE at $t=\tau_1$ disappears, though, for small disorder strength $\Omega=0.1$, LE still shows a tendency of revival. This also
happens for downturn time scale $t=\tau_2$. For completeness, we also checked that for quenching inside the Griffiths phase, the
singular time scale $\tau_1$ and $\tau_2$ completely disappear. We note that the disappearance of the singular time
scale is not artifact of a particular $N$ rather it's originated from the randomness of the transverse field.

In order to explain the disappearance of singularities, we make resort to the quasiparticle picture in the post quench regime. 
For the clean case, $H_{\downdownarrows}$  and $H_{\upuparrows}$ are different from each other with respect to the  local transverse
fields modified  at two sites, where the qubits are coupled. In the language of quasi-particle emission after a quench, one can think 
of two extra separate emitters, located $d$ distance away from each other.  Now, in the symmetric position $d=N/2$, quasi-particles 
need to travel only $d/2$ distance. Hence, there is a constructive interference happening that leads to a partial revival of the initial
state.  Therefore, $\tau_1=d/2v=N/8$ with velocity of the quasiparticle $v=2$. In contrast, for the disordered case, at each point, the 
chain experiences a change in the transverse field due to the finite value of disorder strength $\Omega$; hence, separate emitters, away 
by a distance $d=N/2$, do not really play a role in this global generation of quasiparticles. As an outcome of a very complicated propagation
of quasiparticles from all sites, the constructive interference dies out and the appearance of the singular time scale $\tau_1$ is lost. 
This explanation  remains true for  the extinction of all the singular time scales. This is also true for quenching inside the Griffiths phase.

\section{conclusion}\label{conclusion}
{We study the disentanglement of a  Bell pair constituted of two qubits, which are connected to a disordered 
transverse field Ising spin chain. We identify the disorder induced Griffiths phase from the equilibrium study of LE. 
On the other hand from the non-equilibrium evolution of LE we exhaustively explore decay characteristics of the entanglement 
of the qubits and the non-trivial outcomes associated with the Griffiths phase.} 

{In equilibrium, the derivative of LE with respect to initial transverse field is shown to exhibit a peak at 
critical point of the clean Ising spin chain. The disorder present only  in the transverse field can lead to a local 
ordering in the spin chain; this results in a Griffiths phase \ct{garnerone09} appearing at the junction of ferromagnetic 
and paramagnetic phases. We here numerically estimate the  extent of the Griffith's phase as the derivative of LE exhibits 
a peak near the boundary between Griffith's phase and paramagnetic phase. However, such numerical evaluation of the width 
of the Griffiths phase needs to be verified analytically in further future studies.}

{From the non equilibrium study of the LE following off-critical quenches, we show that the initial 
fall of the LE is completely governed by the coupling strength. Such observation fairly agrees with the earlier analytical 
conjectures in Refs~\ct{sudip_Rossini2007,wendenbaum14}. Our numerical investigation suggests that the quantitative 
nature of the initial fall of LE cannot be explain by a single exponential function. To be precise, a ultra short time 
universal Gaussian fall is observed, followed by two exponential falls with two different decay exponents dependent 
on the coupling strength. Within this time window, the disorder is not able to reflects its effect. Since the late time 
fall of LE depends on the final quenched value of the transverse field~\ct{sudip_Rossini2007}, therefore the asymptomatic 
decay characteristics of LE should be controlled by the disorder strength. The increase in the value of disorder would 
increase the overlap between the initial and the final state of the spin chain. Therefore the initial fast exponential 
fall of LE is substantially suppressed  by the effect of disorder in the long time limit. Consequently we observe a late 
time slow power law fall of LE where the value of decay exponent decreases with the increase of disorder strength. Regarding 
the non-trivial outcome associated with the Griffiths phase, we interestingly observe that  for quenching to the Griffiths phase, 
the late time power law fall is completely absent while initial decay characteristics of LE is similar to that of the for 
the off critical quenching. Hence, the initial exponential decay is related to the initial phase, in contrary the final late 
time power law decay is connected to the final phase of the spin chain. Therefore, the continuous variation of the dynamical 
critical exponent and broadening of relaxation time mark its signature in the late time temporal profile of LE. This unique 
non-equilibrium characteristics of LE for the Griffiths phase might initiate further studies to identify the Griffiths 
phase in a disordered spin systems.}

Moreover, we show that as one increases disorder strength, the singular behavior at  $t=N/8$ and $N/4$ disappear; this is due to the 
fact that quasi-particles, generated from all sites,  interfere destructively and the temporal structure, observed for clean system, 
is washed out. Hence, the linearity between distance, traveled by the quasiparticles, and time, required to travel that distance, breaks 
down for a substantial disorder strength. This indicates that the quasi-particles do not propagate in the light cone like fashion.

\section{Appendix}
The anticommuting  Grassmann  variables are often used  to  compute  physical  quantities for example, expectation values in fermionic
systems with quasifree  fermion  states (i.e., Gaussian states in the form of $c^{\dagger}c$)) \ct{keyl10}. Specifically, the linear  
combinations  of canonically anticommuting Fermi field operators are replaced by linear combinations of complex coefficients that are 
anticommuting  Grassmann  numbers. Based on this ``Grassmann algebra of canonical anticommutation relations", one can calculate the 
expectation value of $\exp(i H_{p}t)\exp(-iH_{s}t)$ with respect to the Gaussian states. The square of fidelity between two states 
$\omega_s$ and $\omega_p$ is given by $L_{p,s}=F(\omega_p,\omega_s)^2$. It has been shown that $L_{p,s}=\omega_s(E_p)$ i.e., the 
expectation value of the support projection $E_P$ in the state $\omega_S$. Using the concept of Gaussian Grassmann integral, one can 
show that 
\begin{equation}
\omega_s(E_p)= \int_Q v(\xi) \exp(\frac{1}{2}\langle \xi^*,(1-P-S)\xi\rangle_Q)
= Pf(1-P-S)
\end{equation}
where $\xi$ is the phase space vector, $P$ and $S$ are two projected covariant operators associated with the states $\omega_p$ and 
$\omega_s$, respectively. $Pf$ represents the pfaffian of a matrix. $v(\xi)$ is the normalization factor, $Q$ indicates  the projection 
to initial ground state. One can use the relation $Pf(A)^2={\rm det}(A)$ to show that $L_{p,s}={\rm det}(1-P-S)^{1/2}$.

Now, connecting to the present case, $\omega_{s,p} \equiv \ket{ \Pi_{\uparrow\uparrow, \downarrow\downarrow}}$ and 
$H_{s,p} \equiv H_{\uparrow \uparrow,\downarrow \downarrow}$. One can consider Eq. (\ref{le1}) of the main text and 
$\omega_{s,p} = e^{-i H_{s,p}t}\lvert\phi_g\rangle $ to arrive at the above connections. We note that $\ket{\phi_g}$, 
being the ground state of initial Hamiltonian $H_I(\Gamma^I)$, is time evolved with the quenched final Hamiltonian 
$H_{\downarrow\downarrow, \uparrow \uparrow}(\Gamma^F, \Delta)$ to obtain the final states $\ket{\Pi_{ \downarrow\downarrow, \uparrow \uparrow}}$.
Similarly, $P (S)$ is the time evolved initial occupation matrix, calculated using transverse field value $\Gamma^I$, 
with Hamiltonian $H_P(\Gamma^F,\Delta) (H_S(\Gamma^F,\Delta))$. The explicit form of the covariant matrix $P(S)$ is 
then given by
\begin{equation}
P(S)=\exp(iH_{p,s} t) {\cal L}(0) \exp(iH_{p,s} t)
\end{equation}
where ${\cal L}(0)$ is the initial occupation matrix composed of $c$ and $c^\dagger$. This can be computed  by diagonalizing 
the initial Hamiltonian $H_I(\Gamma^I)$
as stated in Eq. (\ref{u_mat}).

\begin{acknowledgments}
The authors are grateful to  Bikas K. Chakrabarti for enlightening discussion. TN thanks Kush Saha for critically reading the 
manuscript. 
\end{acknowledgments}

\end{document}